\documentclass[aps,prx,twocolumn,superscriptaddress,nofootinbib,10pt]{revtex4-2}
\usepackage{graphics} 
\usepackage{amsmath} \usepackage{physics} \usepackage{graphicx} 
\usepackage{color} \usepackage{amsfonts,wasysym} \usepackage{amssymb} 
\usepackage{tikz} \usepackage{placeins}
\usepackage{soul}
\newcommand{\bs}[1]{\boldsymbol{#1}} \newcommand{\pauli}[1]{\mathtt{#1}}

\definecolor{LB}{RGB}{134,41,198}

\definecolor{mehran}{rgb}{0.0,0.0,0.8}

\definecolor{ale}{rgb}{0.0,0.5,0.0}

\newcommand{\gs}[1]{\mathrm{GS}(\bs x_{#1})}

\begin{document}

\title{Unsupervised Learning to Recognize Quantum Phases of Matter} 

\author{Mehran Khosrojerdi} \email{mehran.khosrojerdi@unifi.it}
\affiliation{Department of Physics and 
Astronomy, University of Florence, via G. Sansone 1, I-50019 Sesto 
Fiorentino (FI), Italy}

\author{Alessandro Cuccoli} \affiliation{Department of Physics and 
Astronomy, University of Florence, via G. Sansone 1, I-50019 Sesto 
Fiorentino (FI), Italy} \affiliation{ INFN Sezione di Firenze, via G. 
Sansone 1, I-50019, Sesto Fiorentino (FI), Italy } 

\author{Paola Verrucchi} 
\affiliation{ISC-CNR, UOS Dipartimento di Fisica, Università di Firenze, I-50019, Sesto Fiorentino (FI), Italy}
\affiliation{Department of Physics and Astronomy, University of Florence, via
G. Sansone 1, I-50019 Sesto Fiorentino (FI), Italy} 
\affiliation{ INFN Sezione di Firenze, via G. Sansone 1, I-50019, Sesto 
Fiorentino (FI), Italy } 

\author{Leonardo Banchi}
\affiliation{Department of Physics and Astronomy, University of Florence, via
G. Sansone 1, I-50019 Sesto Fiorentino (FI), Italy} 
\affiliation{ INFN Sezione di Firenze, via G. Sansone 1, I-50019, Sesto
Fiorentino (FI), Italy }

\date{\today} 
\begin{abstract}
Drawing the quantum phase diagram of a many-body system in the parameter
space of its Hamiltonian can be seen as a learning problem, which implies labelling the corresponding ground states
according to some classification criterium that defines the phases. 
In this work we adopt unsupervised learning, where the algorithm has no access to any priorly labeled states, 
as a tool for determining
quantum phase diagrams of many-body systems. The algorithm directly works with quantum states: 
given the ground-state configurations for different values of the Hamiltonian
parameters, the process uncovers the
most significant way of grouping them based on a similarity criterion that
refers to the fidelity between quantum states, that can be easily estimated, 
even experimentally.
We benchmark our method with two specific spin-$\frac{1}{2}$ chains, with states 
determined via tensor network techniques. 
We find that unsupervised learning algorithms based on spectral clustering,
combined with ``silhouette'' and ``elbow'' methods for determining the optimal number of 
phases, can accurately reproduce the phase diagrams. 
Our results show how unsupervised learning can
autonomously recognize and possibly unveil novel phases of quantum matter.

\end{abstract}

\maketitle

\section{Introduction}
Quantum many-body systems can exhibit phase-transitions also at 
$T=0$ due to purely quantum fluctuations \cite{sachdev1999quantum}, making 
it possible for Ground States  to feature different collective 
properties for infinitesimally different values of the
hamiltonian parameters. Studying the Quantum Phase Diagrams (QPD) outlined by these 
transitions, as well as the quantum phases themselves, is of interest not only in the 
realm of many-body physics 
but also in quantum information processing,
whose operation happens at temperatures as near as possible to 
$T=0$.
On the other hand,
drawing the QPD of quantum many-body systems is a hard 
task, unless one considers very simple models such as, for instance, the 
Ising chain in a transverse field.
In fact, analytical methods are of limited help in studying models of 
realistic systems, while established numerical techniques demand huge 
computational resources, that become unaffordable as the size of the 
models grows, as required by an evaluation of non-local 
quantities that characterize different phases, such as long-range correlations and  
concurrence, or multipartite entanglement.

In this framework, approaches based on hybrid quantum-classical machine-learning methods provide some alternatives. One possibility comes from quantum-phase classification based on training on labeled data, where some prior theoretical or numerical understanding of the phase-diagram is necessary, at least in a limited parameter region. This approach matches a supervised learning task, where the phases are known for some particular parameter values that are used to train the classification algorithm. The algorithm is then used to explore larger areas of the parameter space and label the corresponding phase \cite{Khosrojerdi_2025, cong2019quantum, monaco2023quantum, huang2022provably, dong2019machine, carleo2019machine, cea2024exploring,uvarov2020machine, carrasquilla2017machine, li2024ensemble,parigi2025supervised}.
Quantum states clustering, instead, uses an unsupervised learning 
strategy, whose goal is to identify similarities among ground states, 
given no prior knowledge of the QPD, and hence without 
pre-existing labeling of the different phases. Unlike classification, 
which is typically applied as a validation tool, clustering can be used 
to explore novel quantum systems, revealing hidden structures in the 
data, and uncovering emergent phases of matter without relying on 
predefined categories \cite{sentis2019unsupervised, chen2021detecting, 
yang2021visualizing,10.21468/SciPostPhys.14.1.005}. Needless to say, clustering quantum states to unveil a phase-diagram is a challenging task, due to the inherently non-local nature of quantum correlations and entanglement \cite{sachdev1999quantum, 
zinn2007phase}: in fact, when the Hilbert space dimension grows, identifying features for an effective clustering  is by itself a difficult task. 
On the other hand, some unsupervised methods \cite{ng2001spectral,von2007tutorial,JMLR:v20:18-170,10.1214/14-AOS1283, 
10.1162/0899766041732396}, even if acting on classical data-sets, can turn out to be suitable to spot similarities between elements of large Hilbert spaces, providing an interpretable framework for clustering quantum states based on their genuinely quantum properties.

Scope of this work is to develop a method for
quantum-states clustering, precise enough to match QPD of 
parameterized Hamiltonians whose ground state can be determined and represented via
tensor networks methods, such as the DMRG \cite{perez2006matrix, orus2014practical, 
ran2020tensor}.
Since our method uses a ``kernel'' between two Hamiltonian parameter sets given by 
the fidelity between   the corresponding ground states, it can  be extended 
to work directly with ground states approximated in quantum devices, 
since the fidelity between states can be experimentally estimated 
using different methods such as the SWAP test or the Hadamard test, 
see e.g.~\cite{schuld2021machine}.

Our procedure goes as follows: be $\hat H(\bs{x})$ a one-dimensional 
Hamiltonian that fulfills the above requirements, with $\bs{x} = (x_1, x_2, \dots x_q)$ 
the set of $q$ parameters upon which it depends. We take four steps
for predicting the QPD with the desired insight into different quantum phases.
\begin{enumerate}
\item Generate $D$ sets of Hamiltonian parameters 
$\{\bs{x}_i=(x_{1i},x_{2i},...x_{iq})\}$
according to some rule (e.g.~uniformly in the parameter space), 
with $i=1,...,D$. This set defines our ``classical data'', where each 
datum is an element of $\mathbb{R}^q$ and a point in the phase-diagram.
\item Represent the Hamiltonian $\hat H(\bs{x}_i)$ as a Matrix Product Operator (MPO) for each datum $\bs x_i$ and find its ground state $\ket{\gs i}$. 
\item 
  Define the \textit{kernel} between 
  two data $\bs x_i$ and $\bs x_j$ as 
  $K_{ij}=|\bra{\gs i}\gs j\rangle|^2$, namely as the fidelity between two ground states $\ket{\gs i}$ and $\ket{\gs j}$.
  In the language of machine learning, this corresponds to defining the \textit{feature space} as the 
  $\mathcal O(4^N)$-dimensional 
  space of density matrices $\ket{\gs i)}\!\!\bra{\gs i}$, where $N$ is the number of qubits in 
  $\hat H(\bs x)$. 
  Note that the dimensionality of the 
  feature space is much larger than that of the dataset, which is one 
  of the key-element of any machine-learning procedure for clustering, as it reflects the richer 
  characterization offered by features w.r.t.~data.
\item Perform a kernel clustering algorithm, which only depends on the kernel
  matrix $K$, so as to group ground states according to the similarity amongst their
  quantum features.  
\end{enumerate}
Notice that, 
although not explicitly considered in this paper, steps 2.~and 3.~can be replaced with: 
\begin{enumerate}
  \item[2b.] Approximate each ground state $\ket{\gs i}$ in a quantum device, 
    e.g.~using a parametric quantum algorithm and the variational quantum eigensolver \cite{tilly2022variational} 
     or more recent versions \cite{grimsley2019adaptive}.
  \item[3b.] Compute the kernel $K_{ij}=|\bra{\gs i}\gs j\rangle|^2$ using the quantum hardware,
    e.g.~via SWAP test  or alternative methods \cite{schuld2021supervised}. 
\end{enumerate}
In any case, possible approximations must be considered.
With tensor networks the approximation follows from the finite value of the bond dimension, which we denote 
by $\chi$. On the other hand, when performing 
experiments in a quantum hardware, the approximations follow from  the poor convergence of the variational algorithm,  the imperfect estimation of the kernel (due to shot noise), and the imperfect quantum gates and measurements. 
Be that as it may, the end effect is a noisy estimate of the kernel matrix. 
In this paper, we deliberately use tensor networks with small $\chi$, so as to assess the performances of our clustering technique under imperfect data.

\begin{figure}[t]
    \begin{center} {\includegraphics[width=0.45\textwidth]{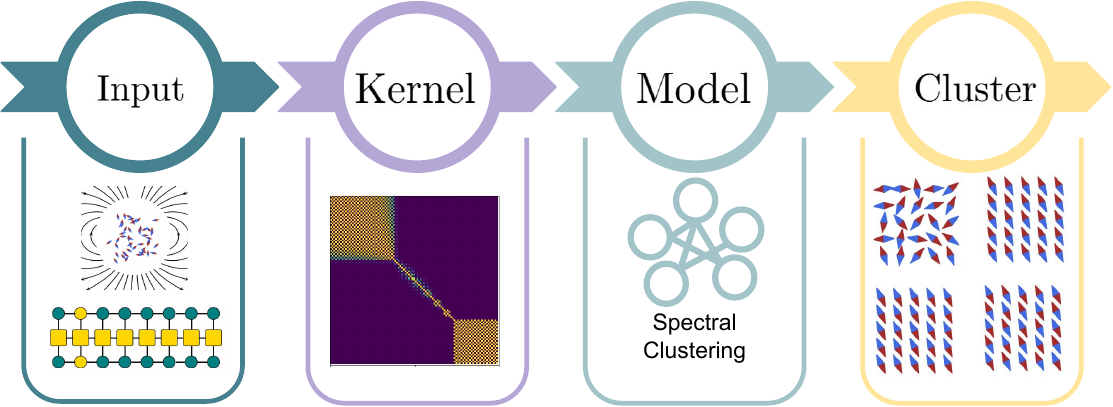}} 
    \end{center} 
    \caption{Outline of the clustering procedure for quantum ground states, as described in the introduction. 
    The process begins with randomly producing ground states of a target Hamiltonian, either using classical approximations (e.g.~tensor networks), a quantum hardware with physical spins, or a quantum computer. 
    The kernel is then estimated either in hardware or via tensor network techniques. 
    Next, a kernel-driven clustering model groups the ground states based on their correlation structure, capturing quantum similarities, including entanglement properties. Finally, we present the resulting clusters.} 
    \label{fig:outline}
\end{figure}

A schematic view of the overall process is provided in Fig.~\ref{fig:outline}.
In order to benchmark the reliability of the obtained 
QPD we will demand that, within the domain where DMRG remains 	
accurate, there exists at least one subregion of the parameter-space in 
which the phase to which the ground state belongs is already known and possibly 
characterized.

The remainder of this article is organized as follows. Section~\ref{sec:classical_quantum} introduces the dataset construction. Section~\ref{sec:clustering} describes the kernel-based unsupervised model and the mathematical methods used to determine the optimal number of clusters. Section~\ref{sec:results} presents benchmark studies on the Axial Next-Nearest-Neighbor Ising (ANNNI) and Cluster-Ising models. Finally, Section~\ref{sec:conclusion} summarizes the findings and highlights the key contributions of this work.

\section{From classical data to quantum features\label{sec:classical_quantum}}
\label{s.c2q}
We consider one-dimensional 
spin-$\frac{1}{2}$ models, with short enough interaction-range 
for DMRG to be efficient in numerically determine their ground state
\cite{Khosrojerdi_2025,arizmendi1991phase,PhysRevB.76.094410,PhysRevE.75.021105,Nagy_2011,cong2019quantum,caro2023out,PhysRevA.2.1075,mbeng2024quantum}; 
these models often exhibit nontrivial QPD that we 
explore by moving around in their parameter space, which is $\mathbb{R}^q$.
As explained in the introduction, each point in this space is a datum
$\bs{x}_i=(x_{i1},x_{i2}...,x_{iq})$, in one-to-one correspondence with the Hamiltonian $\hat 
H(\bs{x}_i)$ and its ground state $\ket{GS(\bs{x}_i)}$.
We wander erratically in this parameter-space, randomly generating $D$ data $\{\bs{x}_i\}, 
i=1,...,D$, assuming no prior knowledge about the different phases, let 
alone the QPD.

Specifically, we will consider model-Hamiltonians $\hat H$ that can be decomposed as $\hat{H} = 
\sum_{j=1}^N \hat{H}_{j}$, where each local Hamiltonian $\hat{H}_{j}$ involves a small number of sites near $j$, thus acting non-trivially only on the subspace of a limited number of spins, so that a compact and computationally efficient representation in terms of MPO 
is possible. In fact, for Hamiltonians defined on an open chain, it can be shown that
\begin{equation}
	\hat{H} = \hat \ell_1 \hat{W}_2 
  \hat{W}_3\cdots\hat{W}_{N-1} \hat r_N, 
	\label{eq:H}
\end{equation} 
where each matrix element $(\hat W_j)_{lm}$ is an operator acting on the $j$th spin Hilbert space and the boundary terms are such that 
$\hat \ell_j$ and $\hat r_j$ are the first row and the last column of the $\hat W_j$ matrix, respectively. The multiplication in Eq.~\eqref{eq:H} corresponds to a matrix product in the representation space and a tensor product in the physical one 
\cite {parker2020local}; the specific structure of the operator-matrix $\hat W$ is 
not identifiable at first sight, but can be obtained with the help of a graphical representation 
borrowed from finite-state automata, as we explain below. 

\begin{figure}[t]
	\centering
	\includegraphics[width=0.40\textwidth]{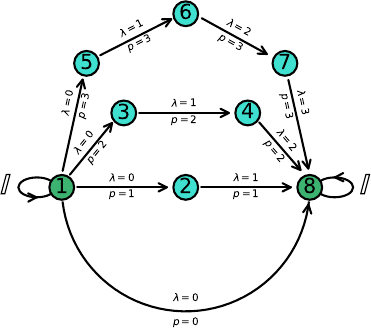}  
	\caption{Graphical representation of a typical local Hamiltonian, illustrating how the automaton structure translates into the explicit MPO representation.} 
	\label{fig:LHtypical}
\end{figure}

Be $d$ the distance (expressed in number of lattice sites) 
between the furthest interacting spins: terms in the Hamiltonian involve at most $d+1$ single-spin operators. We assume
that the local Hamiltonian $\hat H_j$ can be written as 
\begin{equation} 
\hat H_j=\sum_{p=0}^d \bigotimes_{\lambda=0}^p~_\lambda^p\hat h_{j+\lambda}, 
\label{eq:H_J} 
\end{equation} 
where the index in brackets indicates that the single-site operator 
$^p_\lambda\hat h_{(j+\lambda)}$ acts upon the system sitting at site $j+\lambda$. For instance, 
if the model includes next to next-nearest neighbours interactions, it is $d=3$ and we can expect tensor products of at most $d+1=4$ single-spin operators in each $\hat H_j$. A graph of numbered nodes and labelled links can be drawn from Eq.~\eqref{eq:H_J} as shown in Fig.(\ref{fig:LHtypical}): 
$d+1=4$ paths spout from node $1$, each path formed by an increasing number of links, 
from $1$ to $d+1=4$, and each ending in the last node, numbered $d(d+1)/2+2=8$. Different paths stand
for different types of interaction, with 
the number of links equal to the number of single-site spin operators involved. Referring to 
Eq.~\eqref{eq:H_J} this means that paths are labelled by the index $p$. Paths with increasing 
number of links are drawn from bottom to top so that the lowest refers to local 
terms, the second one to nearest-neighbours interaction, the third one to next-nearest neighbours 
interaction, and so on. Links are labeled by the pair ($p$,$\lambda$), with $p$ identifying 
the path to which they 
belong, and $\lambda$ counting their position w.r.t. the first node, so that the first link of 
each path has $\lambda=0$, the second one 
$\lambda=1$ and so on, until the $(d+1)(d+2)/2=10$th link, having $\lambda=d=3$ (that only 
appears in the $d+1=4$th path). 
Finally, nodes are numbered in increasing order from bottom to top and left to right. 

Once the graph is drawn, the operator-matrices in Eq.~\eqref{eq:H} have 
dimension equal to the number of nodes, i.e. $d(d+1)/2+2=8$, with elements $(\hat W_j)_{lm}$ different 
from zero if there exists a link $(p,\lambda)$ connecting nodes $l$ and $m$, in which case 
it is $(\hat W_j)_{lm}=^p_\lambda\hat h_j$. 
Specific examples of such graphs, together with their corresponding operator matrices, are shown in Fig.~\ref{fig:LHtypical} and Equation \ref{eq:typical_H_2}, respectively
\begin{align}
    \hat{W}_j = 
    \begin{pmatrix} 
    \mathbb{I} & _0^1\hat{h}_j &  _0^2 \hat{h}_j & 0 & _0^3 \hat{h}_j & 0 & 0 & _0^0 \hat{h}_j\\ 
    0 & 0 & 0 & 0 & 0 & 0 & 0 & _1^1 \hat{h}_j\\ 
    0 & 0 & 0 & _1^2 \hat{h}_j & 0 & 0 & 0 & 0\\ 
    0 & 0 & 0 & 0 & 0 & 0 & 0 & _2^2 \hat{h}_j\\ 
    0 & 0 & 0 & 0 & 0 & _1^3\hat{h}_j & 0 & 0\\ 
    0 & 0 & 0 & 0 & 0 & 0 & _2^3 \hat{h}_j & 0\\
    0 & 0 & 0 & 0 & 0 & 0 & 0 & _3^3\hat{h}_j\\ 
    0 & 0 & 0 & 0 & 0 & 0 & 0 & \mathbb{I}\\ 
    \end{pmatrix}\,.
    \label{eq:typical_H_2}
\end{align} 

The graphical representation encodes the same operator structure as the matrix in Eq.~\eqref{eq:typical_H_2}, but in a more compact and intuitive form. Each edge (green circles) in the graph corresponds to a possible transition weighted by a local operator, while the adjacency-like structure (blue circles) directly translates into the block entries of the operator matrix. In this way, the finite–state automaton view and the matrix formulation provide two complementary perspectives of the same MPO construction.

MPO representations allow for an adequate implementation of variational algorithms such as 
the DMRG, whose numerical efficiency relies on the compactness of such representation.
Within the algorithm, an initial state 
is iteratively modified via a unitary 
process that depends on the parameters $(\bs{x}_i)$ so as to minimize its energy.
Without entering into the details of this well established method, 
for which we refer the reader to Ref.~\cite{schollwock2011density}, we here mention 
the relevance of the bond dimension $\chi$. Its logarithm gauges 
the number of spins that we expect to be significantly entangled in the ground state of our chain,
not to be mistaken for the range of the interaction or the size of the $\hat W_j$ matrices. Could 
one numerically deal with a bond dimension as large as $\mathcal O(2^N$), then the algorithm 
would be exact. 
However, as this is not 
possible for 
large $N$, the algorithm involves a well-thought-out adjustment of the bond dimension.
 In our case, for instance, we ask for 
{\it i)} a reasonable computational time  for producing as many ground states as necessary for feeding our unsupervised 
classification, 
and {\it ii)} a knowledge of their structure precise enough to obtain a QPD that makes sense and 
can be understood in terms of qualitative features of ground states in different phases.

\section{Clustering}\label{sec:clustering}

Let us now focus on clustering approaches:
these are all aimed at grouping objects according to some sort of similarity  principle, that can 
be the most diverse depending on the specific goal of the clustering itself.
The key concept of the process is that of {\it affinity} betweeen two 
data, as measured by some quantity whose domain is not the space where the data are defined, 
but rather the so-called feature space.
This quantity must be defined for all possible pairs of the dataset upon which the clustering is 
due, and then organized in what is usually called affinity matrix, or {\it  kernel}.

We choose to measure the pairwise affinity between any two data $\bs{x}_i,\bs{x}_j$ via the 
square fidelity between the corresponding ground states, as obtained by MPO and DMRG,
\begin{equation}
  K_{ij}= K(\bs{x}_i, \bs{x}_j) = |\bra{\gs i}\gs j\rangle|^2~,
\label{eq:K_ij}
\end{equation}
organized in the $D\times D$ real positive matrix, that is our Kernel.
Notice that the distance between 
$\bs{x}_i$ and $\bs{x}_j$ in $\mathbb{R}^q$ has nothing to do with the actual value of 
$K_{ij}$, which is rather related with the distance  between 
$\ket{\gs i}$ and 
$\ket{\gs j}$ as induced by the inner-product in the $2^N$-dimensional Hilbert space 
of our chain.

Once the Kernel is numerically obtained, we can proceed with the clustering process; amongst the 
various techniques available for the scope, spectral clustering via k-means algorithm emerges as a 
suitable choice, and this is how it goes: 
first a ``similarity'' graph is generated from the quantum Kernel, in such a way that indeces $i,j$ of its matrix-elements identify nodes $i$ and $j$ of the graph, and edges are drawn 
if $K_{ij}$ is larger than a threshold value $\tau <1$ chosen at will;
the graph Laplacian $L$ is then determined, via $L_{ij}=A_{ij}-D_{ij}$, with 
 $A_{ij}=1$ if $i$ and $j$ are connected and
zero otherwise, while $D_{ij}=\delta_{ij} d_i$ where $d_i$ is the degree of the $i$th node, 
i.e.~the number of nodes with which it is connected ($A_{ij}$ and $D_{ij}$ are the elements of the so 
called adjacency and degree matrices, respectively).
Once the eigendecomposition of $L$ is performed, providing $D$ eigenvalues and eigenvectors, a 
number $c<D$ is chosen, to be the desired number of clusters into which grouping the data. Then, 
the $c$ eigenvectors corresponding to the $c$ largest eigenvalues are 
retained, to become the coloumns of a  $D\times c$ matrix, $\Delta$, whose rows are the 
$D$ elements to be finally clustered, while elements of each row are the $c$ features to be used 
for clustering. Notice that each row $\Delta_{i*}$ is in bijective  relation with a datum 
$\bs{x}_i$ via the row-index $i$, meaning that grouping the rows effectively amount to cluster 
the original data.

The sheer clustering is then performed via the standard k-means algorithm, whose goal is 
that of grouping the $D$ rows of the above matrix $\Delta$ into $c$ different 
clusters, based on the $c$ features selected when ignoring all but the $c$ largest 
eigenvalues of the Laplacian.
Reasons for performing the spectral analysis of the Laplacian rather then the Kernel itself 
reside in the particular way $L$ retains properties of the Kernel that are relevant for 
clustering (whether $K(i,j)$ is larger than $\tau$ or not), ignoring useless information (the specific 
value of each $K(i,j)$).

Without entering the details of the k-means algorithm, for which we refer the reader to the 
literature \cite{ikotun2023k}, 
we recall that $c$ is a crucial and yet arbitrary number and ask ourselves how to 
properly choose it. 
In fact, if $c$ is too near to $D$ we have an essentially useless clustering, with just a few 
data in each cluster and no significant computational advantage.
On the other hand, too low a $c$ might imply the loss of resolution and information in our 
classification process, computationally fast but in vain.

There are several ways to determine the optimal number of clusters $c$ in k-means 
clustering, amongst which the \textit{Elbow Method} \cite{shi2021quantitative,onumanyi2022autoelbow} and the \textit{Silhouette method} \cite{rousseeuw1987silhouettes}, that belong to different classes of approaches, often referred to as within-cluster and between-clusters, respectively, to represent their different logic.
In fact  Elbow Method provides for plotting the Within-Cluster Sum of Squares (WCSS), a distance metric that measures the spread of data points within each cluster, as a function of $c$, so as to find where WCSS stops being strongly dependent on $c$ itself, which is where the ``elbow" is placed in the WCSS curve that reminds of a bended arm as $c$ increases.
After the ``elbow'' point further increases in $c$ lead to marginal reductions in WCSS, indicating that adding more clusters provides little additional value and may even result in overfitting \cite{bholowalia2014ebk, kodinariya2013review}. 
Unfortunately, finding the elbow point is not enough, as some sort of validation is necessary to avoid dealing with a truly optimal $c$ and not with the result of some arbitrary fitting. 
To this aim we consider the \textit{Silhouette Method}, that assesses the quality of a $c$-clustering by evaluating the so called silhouette width 
\begin{equation}
    s(i) = \frac{b(i) - a(i)}{\max(a(i), b(i))} ,
    \label{eq:silhouette_formula}
\end{equation} 
where $i$ refers to one of the elements to be clustered, $a(i)$ is the average distance from $i$ to all other elements in the same cluster, and $b(i)$ is the minimum average distance from $i$ to elements in other clusters, with the distance being defined by the same metric employed in the clustering procedure.
In this paper all distances are evaluated starting from the quantum kernel matrix. 

The silhouette width ranges from $-1$ to $1$, with values close to $1$ indicating well-clustered elements, and nearly-zero values suggesting that the element could equally well belong to another cluster. The average silhouette width across all elements is used to determine the best $c$; the largest average silhouette width indicates the most suitable number of clusters. This method is particularly effective as it balances within-cluster tightness and separation from other clusters, providing a clear metric for evaluating clustering performance \cite{ROUSSEEUW198753,MONSHIZADEH2022103513}. Ultimately, a high silhouette score for the value of $c$ in the elbow region confirms that the clusters are well-separated, thereby validating the selection of the elbow point as an appropriate choice. 

An effective way of visualizing the quality of the clustering as obtained for a certain value of $c$ is the so called silhouette plot,
examples of which are shown in Figs.~\ref{fig:annni}(b) and \ref{fig:topo}(b). The plot is drawn as follows:
to each element to be clustered, say $\bs{x}_i$, is associated a horizontal bar 
whose length is proportional to the silhouette score $s(i)$ defined in Eq.~\eqref{eq:silhouette_formula}. 

Negative values are shown as bars to the left of the vertical axis.
Bars are grouped in bands according to the cluster 
assigned to their respective elements and piled bottom-up for increasing $s(i)$ within each band. 
The order in which different bands are displayed is arbitrary.
The vertical width of each band visualizes the number of elements assigned to the respective 
cluster by the spectral algorithm.
A vertical line indicating the average silhouette score is also shown: a band with a large number 
of bars exceeding such average indicates a well identified cluster, while bars to the left 
suggests the corresponding elements have been assigned to the wrong cluster.

\section{Results\label{sec:results}} 
\subsection{Axial Next-Nearest-Neighbor Ising (ANNNI) 
Model}

Our first case study is the  Axial 
Next-Nearest Neighbor Ising (ANNNI) chain of $N$ spin-$\frac{1}{2}$ particles
\cite{elliott1961phenomenological, fisher1980infinitely, 
arizmendi1991phase, fumani2021quantum}, defined by the Hamiltonian 
\begin{equation}
	H = J\left[-\sum_{j=1}^{N-1}\sigma_j^{\pauli x}\sigma_{j+1}^{\pauli x} 
	+ k \sum_{j=1}^{N-2}\sigma_j^{\pauli x}\sigma_{j+2}^{\pauli x} 
	-h \sum_{j=1}^N \sigma_j^{\pauli z}\right]~,\label{eq:annni}
\end{equation} 
where $\sigma^{\pauli x}_{j}$ and $\sigma^{\pauli z}_{j}$ 
are the Pauli operators acting on the spin at site $j$.
We set the overall energy scale $J=1$ and take $k$ and $h$ positive so that, besides the ferromagnetic nearest-neighbour (NN) exchange along the $x$ direction, described by the first term in 
Eq.~\eqref{eq:annni}, there is an antiferromagnetic interaction between next-nearest-neighbours (NNN), gauged by the value of  $k$,  and a uniform magnetic field $h$ pointing in the positive $z$ direction. When $h = 0$ the system behaves classically, as all terms in the Hamiltonian commute with each other. When $k = 0$ the model reduces to the celebrated Ising model in a transverse field, exactly solvable via fermionization and extensively studied since the '70th of the last century -- see Ref.~\cite{mbeng2024quantum} and references therein for a review.
For $k > 0$ the system exhibits frustration, as the first term tends to align all spins along the $x$-axis while the second one promotes antialignment of NNN. 
The phase diagram of the ANNNI model has been explored using various analytical and 
numerical approaches \cite{arizmendi1991phase, fumani2021quantum, beccaria2007evidence, chandra2007floating, suzuki2012quantum, dutta2015quantum}, and the possible use of tensor network techniques to determine its ground state has been investigated in Refs.~\cite{Khosrojerdi_2025, 10.21468/SciPostPhys.14.1.005, beccaria2007evidence, nagy2011exploring}. 

For small values of $h$ and $k$, the NN interaction is the dominant term, leading to a ferromagnetic phase. As $h$ increases beyond a critical threshold the system enters a paramagnetic phase where all spins get a positive $z$-component while pointing in random directions on the $xy$ plane. When $k$ is large and $h$ stays sufficiently small the NNN interaction takes over, stabilizing the so-called antiphase order, in which spins align along $x$ in pairs, with alternate sign from one pair to the other. Between the antiphase and paramagnetic phases, the system enters a floating phase and, in the thermodynamic limit $ N \to \infty $, becomes gapless, indicating the emergence of a critical behavior. 

\begin{figure*}[t] \centering 
	\includegraphics[width=0.325\textwidth]{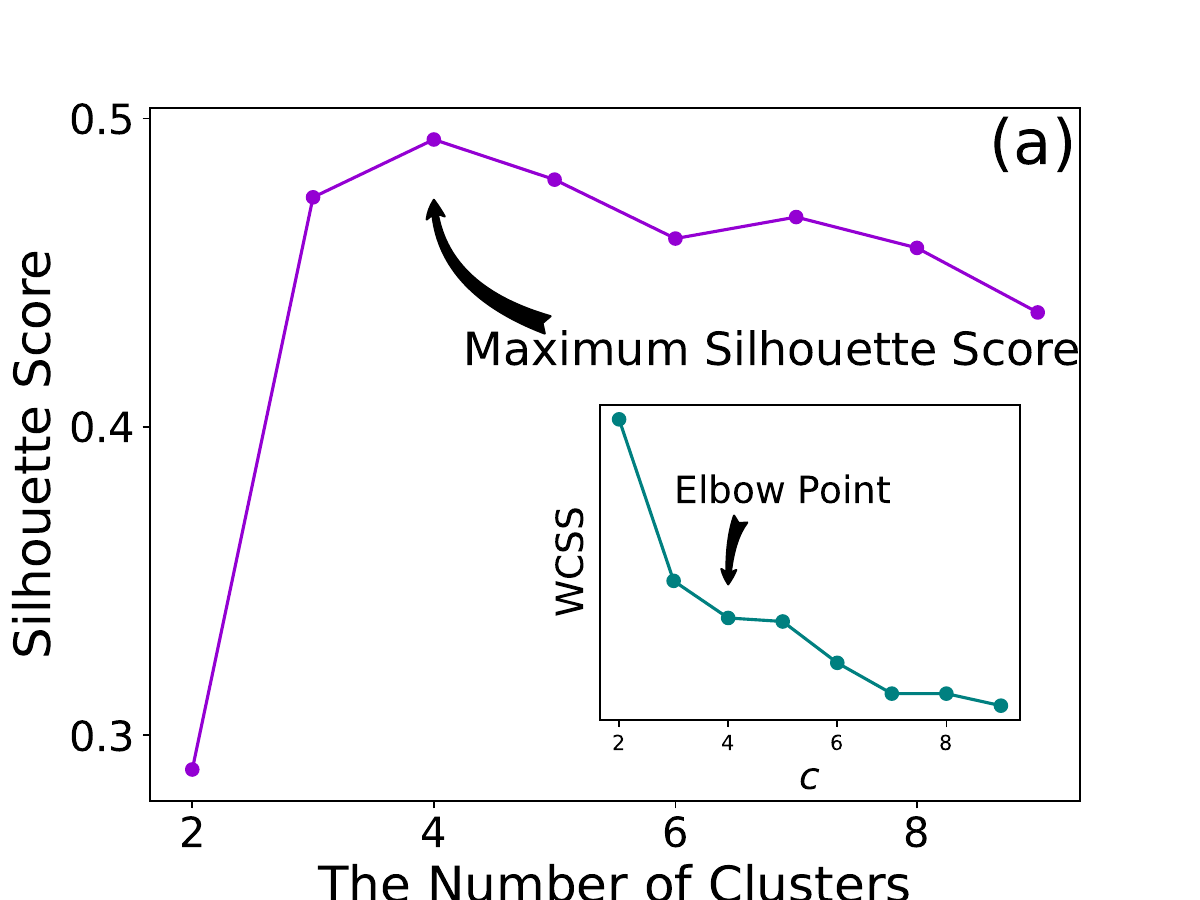} 
	\includegraphics[width=0.325\textwidth]{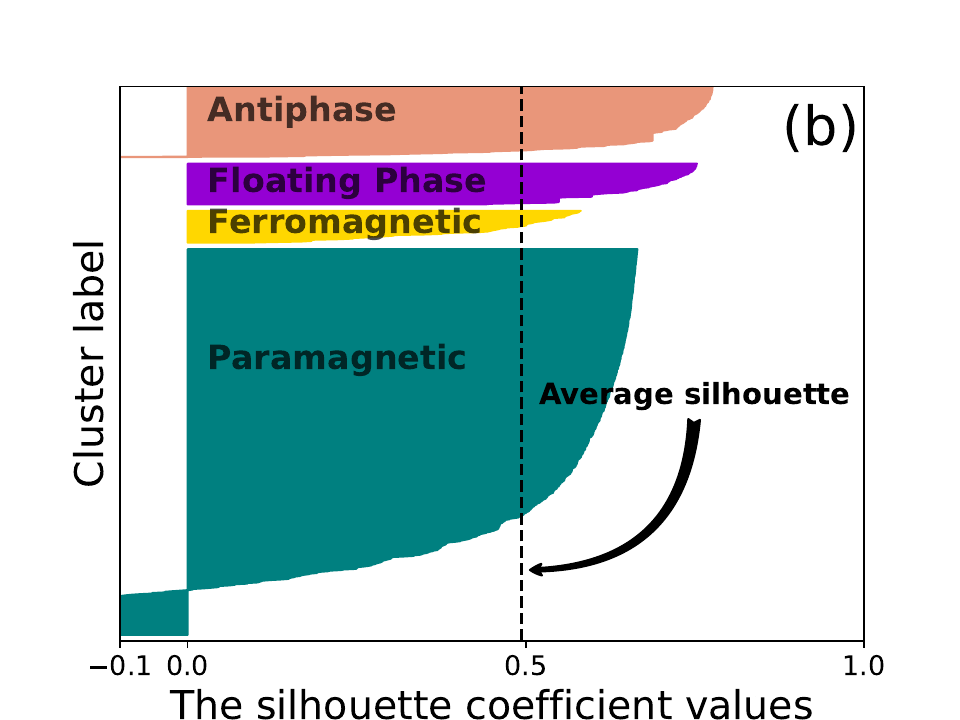}
	\includegraphics[width=0.325\textwidth]{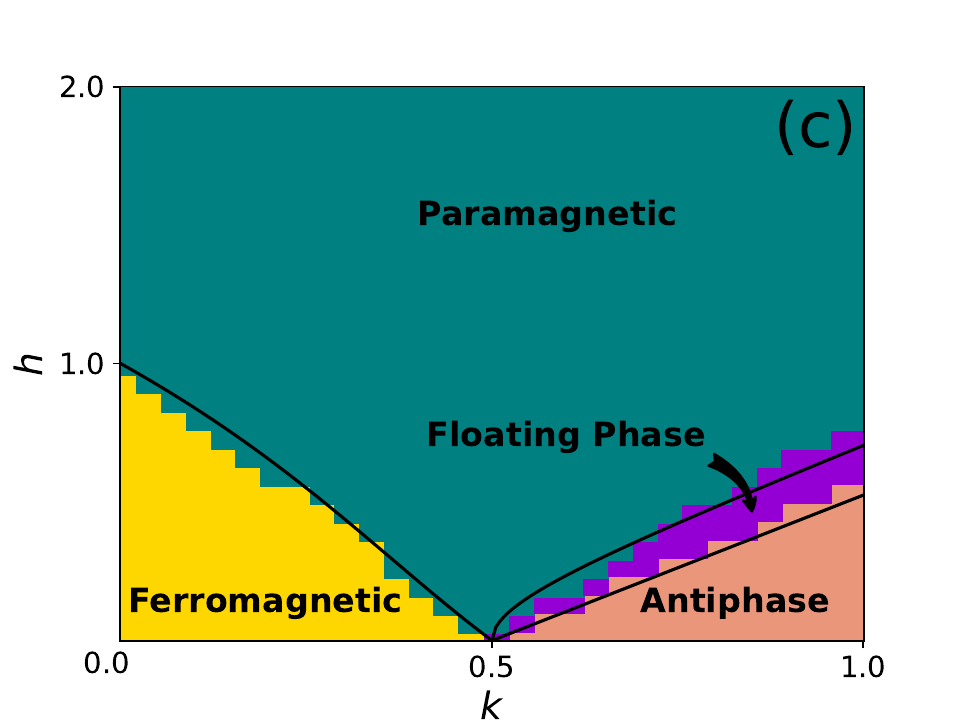} 
	\caption{ Results for the ANNNI Hamiltonian \eqref{eq:annni} with $N = 51$ spins, where 
	the ground states are simulated with a bond dimension $\chi = 20$. 
    (a) Average silhouette score as a 
    function of the number of clusters $c$; the inset 
    shows the elbow point, identified by a significant change 
    in the WCSS dependence on $c$. (b) Silhouette plot for $c=4$, as determined by Fig.~\ref{fig:annni}a. 
  (c)~The predicted phase diagram.
} \label{fig:annni}
\end{figure*}

Although the general structure of the QPD is well established, there persist discrepancies in the identification of phase boundaries as obtained by different methods \cite{bonfim2017quantum, nemati2020comment}. In particular, while the primary phases depicted in Fig.~\ref{fig:annni}c are widely recognized, some studies propose that the paramagnetic phase may split into two distinct regions \cite{fumani2021quantum}, while others suggest the existence of an extended floating phase, with evidence reported in various studies~\cite{arizmendi1991phase,sen1992numerical,bonfim2017quantum}. In Fig.~\ref{fig:annni} (c) we mark with black lines the phase transitions as previously obtained by perturbative and numerical techniques \cite{beccaria2007evidence,cea2024exploring}.

\begin{figure}[t]
	\begin{center} {\includegraphics[width=0.40\textwidth]{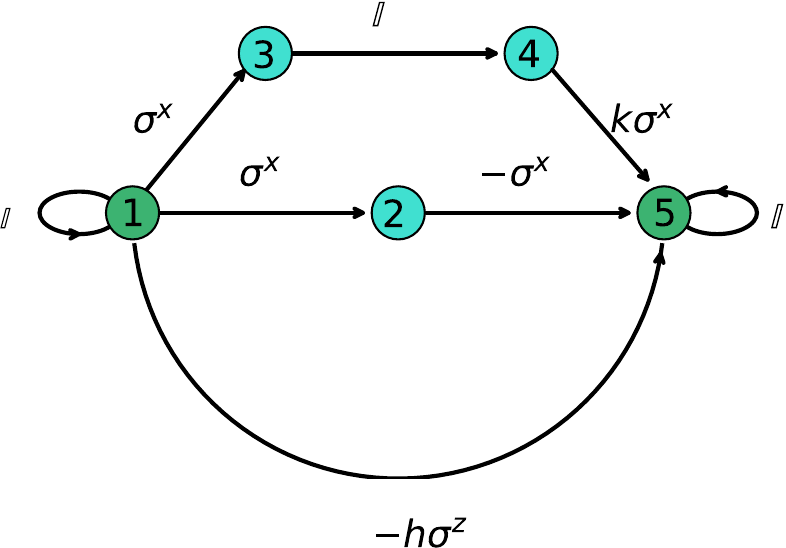}} \end{center} 
  \caption{Local interaction in the MPO formalism for ANNNI model using finite-state automata graphichal representation. }
	\label{fig:LHANNNI}
\end{figure}

As for this work, referring to Eq.~\eqref{eq:H_J} the ANNNI hamiltonian \eqref{eq:annni} has $d=2$, $^0_0h_{j}=h\sigma ^z_j$, $\otimes_{\lambda=0}^1\,^1_\lambda h_{j+\lambda}=-\sigma^x_j\sigma^x_{j+1}$, and $\otimes_{\lambda=0}^2 h_{j+\lambda}=\,k\sigma^x_j\sigma^x_{j+2}$;
using the prescription given in Sec.~\ref{s.c2q} we draw the corresponding graph 
(see Fig.~\ref{fig:LHANNNI}) to which it corresponds the MPO
\begin{align}
    \hat{W} = \begin{pmatrix} \mathbb{I} & \sigma^{\pauli x} & 
    \sigma^{\pauli x} & 0 & -h \sigma ^{\pauli z}\\ 0 & 0 & 0 & 0 & 
    -\sigma^{\pauli x}\\ 0 & 0 & 0 & \mathbb{I} & 0\\ 0 & 0 & 0 & 0 & k 
    \sigma^{\pauli x}\\ 0 & 0 & 0 & 0 & \mathbb{I}\\ 
    \end{pmatrix}. 
    \label{eq:canonical_mps_ANNNI}
\end{align} 

Our results for a chain of $N = 51$ spins, sampling $30\times30$ pairs $(k,h)$, are shown in Fig.~\ref{fig:annni}. The ground states were obtained using the DMRG algorithm, as implemented in the \verb+quimb+ 
library \cite{gray2018quimb}. In order to choose the optimal $c$ for the spectral clustering  we have used the silhouette method: Fig.~\ref{fig:annni}a shows $s(i)$ as a function of the number of clusters $c$: the largest value is seen for $c=4$, suggesting that four clusters might provide a well-separated grouping of the data.
Moreover, the elbow method applied to WCSS is shown in the inset: an evident 
inflection point is observed at  $c=4$.
Fig.~\ref{fig:annni}b shows the silhouette-plot for $c=4$, with the dashed vertical line marking the average silhouette score: the plot confirms the effectiveness of spectral clustering with  $c=4$, though some elements in 
the paramagnetic phase have negative $s(i)$, indicating misclustering. The clusters
corresponding to the  ferromagnetic and floating phases have smaller but entirely positive  silhouette scores, suggesting well-defined boundaries. The antiphase and the paramagnetic phase, though, despite having relatively high silhouette scores, show variations in width, pointing to fluctuations in phase boundaries. 

Fig.~\ref{fig:annni}c shows our QPD and the previously predicted boundaries \cite{beccaria2007evidence,cea2024exploring},
indicated by the black lines. Discrepancies are seen near the phase transitions, 
which may be attributed either to finite-size effects, 
or to the DMRG  bond dimensionm $\chi$ that we have set to 20. In fact, near 
transitions the entanglement area law breaks down 
\cite{eisert2010colloquium}, potentially requiring a larger bond 
dimension to accurately capture the correlations of the system.

\subsection{Cluster-Ising Model with a symmetry-protected topological phase}

Amongst the many different phases possibly featured by QPD of many-body systems,
the so called symmetry-protected topological (SPT) ones are particularly intriguing.
An example of such phases can be found in the renowned Haldane chain, where 
the protecting symmetry is the discrete $\mathbb{Z}_{2} \times \mathbb{Z}_{2}$ one. This same symmetry, realized via $\pi$ rotations around the $x$-axis of all spins along the chain, is also featured by the spin-$\frac{1}{2}$ chain with hamiltonian
\begin{equation}
	H = J\left[-\sum_{j=1}^{N-2}\sigma_j^{\pauli z}\sigma_{j+1}^{\pauli 
	x}\sigma_{j+2}^{\pauli z} - h_2 \sum_{j=1}^{N-1}\sigma_j^{\pauli 
	x}\sigma_{j+1}^{\pauli x} -h_1 \sum_{j=1}^N \sigma_j^{\pauli x}\right]~, 
	\label{eq:Htopo}
\end{equation}
{where the first term encodes the 1D cluster interaction, $h_2$ represents an Ising coupling, $h_1$ is a transverse field, and $J>0$ sets the overall energy scale, whose value does not enter the QPD. For dominant cluster interaction, i.e.~small $h_1$ and $h_2$, and in the presence of the protecting $\mathbb{Z}_{2}\times\mathbb{Z}_{2}$ symmetry, the ground state of open chains realizes a short-range entangled, gapped SPT phase characterized by spin-$1/2$ edge modes, nonlocal string order, and twofold entanglement-spectrum degeneracy \cite{ohta2016topological, son2011quantum, smacchia2011statistical, skrovseth2009phase}. These features remain robust as long as the system is gapped; they disappear only when the bulk gap closes at a phase transition \cite{son2011quantum, smacchia2011statistical,strinati2017resilience}. Competing with this cluster phase are the paramagnetic and antiferromagnetic phases, reached respectively by increasing $h_1$ or $h_2$, with critical behavior that departs from the conventional Ising universality class.
Recent studies have further evidenced the presence of this SPT phase through quantum convolutional neural network circuit~\cite{cong2019quantum, herrmann2022realizing} and by a supervised learning approach~\cite{Khosrojerdi_2025}.

In what follows we put our method to the test by applying it to 
the model \eqref{eq:Htopo}, with $J=1$ and $h_{1,2}$ free to take positive or negative values.
Referring to Eq.~\eqref{eq:H_J}, it is $d=2$,
$^0_0h_{j}=-h_1\sigma^x_j$, $\otimes_{\lambda=0}^1\,^1_\lambda 
h_{j+\lambda}=-h_2\sigma^x_j\sigma^x_{j+1}$, and 
$\otimes_{\lambda=0}^2\,^2_\lambda h_{j+\lambda}=-\sigma^z_j\sigma^x_{j+1}\sigma^z_{j+2}$;
using the prescription given in Sec.~\ref{s.c2q} we draw the corresponding graph, 
see Fig.~\ref{fig:LHtopo}, to which it corresponds the MPO
\begin{align}
    \hat{W} = \begin{pmatrix} \mathbb{I} & \sigma^{\pauli x} & 
    \sigma^{\pauli z} & 0 & -h_1 \sigma ^{\pauli x}\\ 0 & 0 & 0 & 0 & 
    -h_2\sigma^{\pauli x}\\ 0 & 0 & 0 & \sigma^{\pauli x} & 0\\ 0 & 0 & 
    0 & 0 & - \sigma^{\pauli z}\\ 0 & 0 & 0 & 0 & \mathbb{I}\\ 
    \end{pmatrix}. 
    \label{eq:canonical_mps_Haldane}
\end{align} 

\begin{figure}[t] \begin{center} 
    {\includegraphics[width=0.40\textwidth]{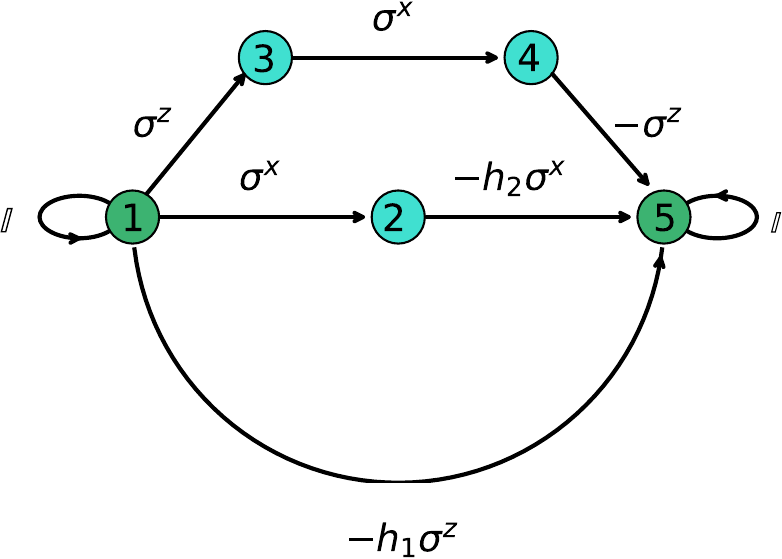}} \end{center} 
    \caption{Local interaction in the MPO formalism for Cluster-Ising model using finite-state automata graphichal representation.}
    \label{fig:LHtopo}
\end{figure}

\begin{figure*}[t] \centering 
    \includegraphics[width=0.325\textwidth]{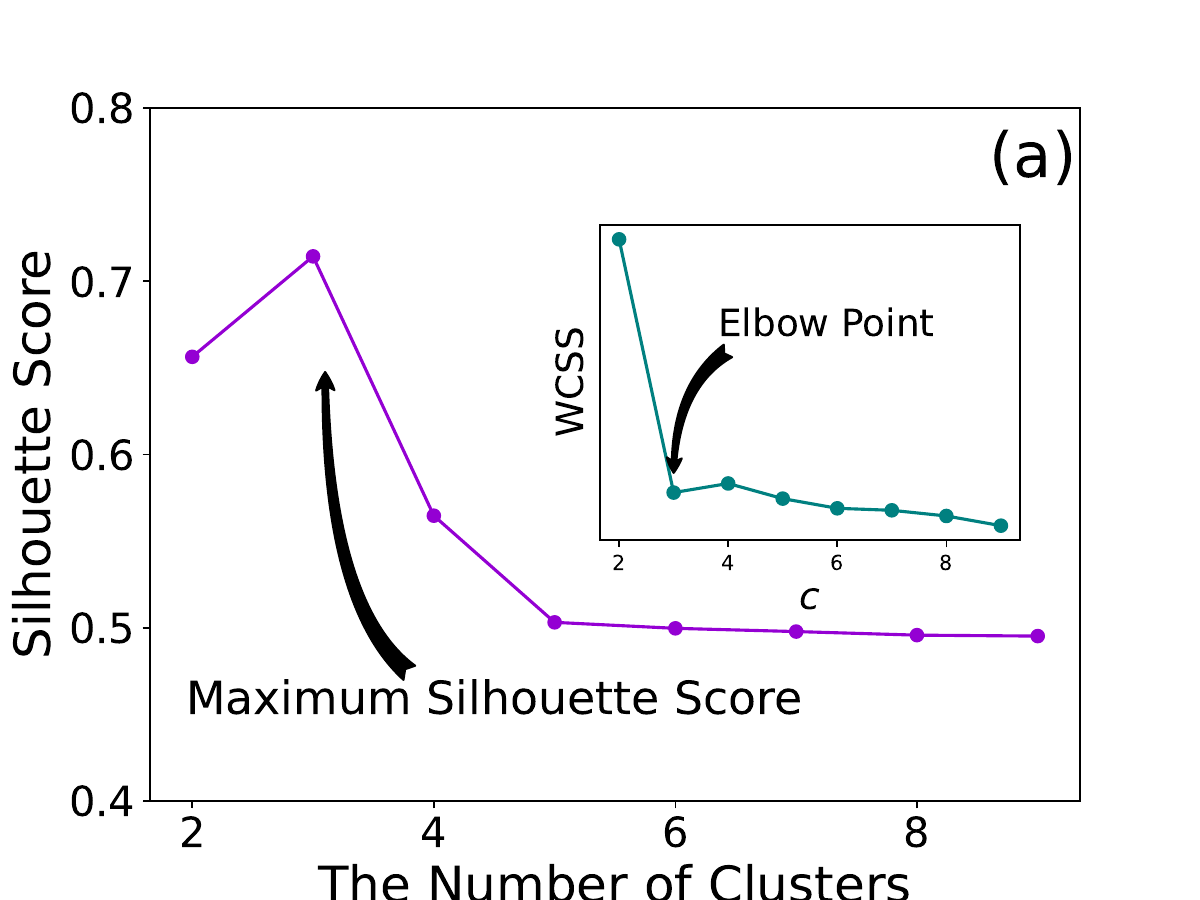} 
    \includegraphics[width=0.325\textwidth]{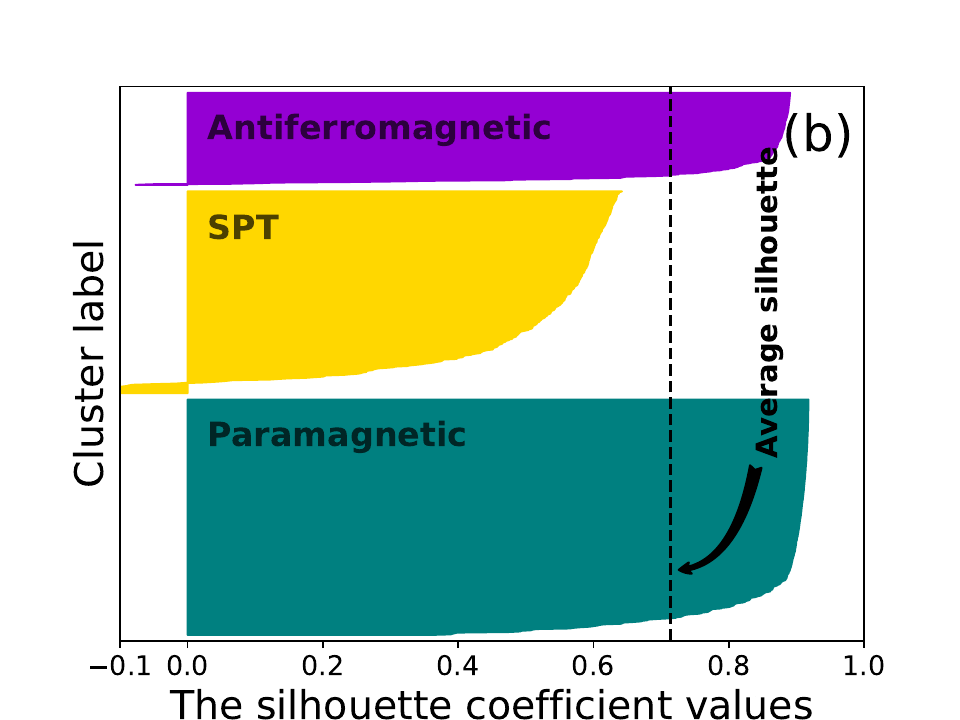} 
    \includegraphics[width=0.325\textwidth]{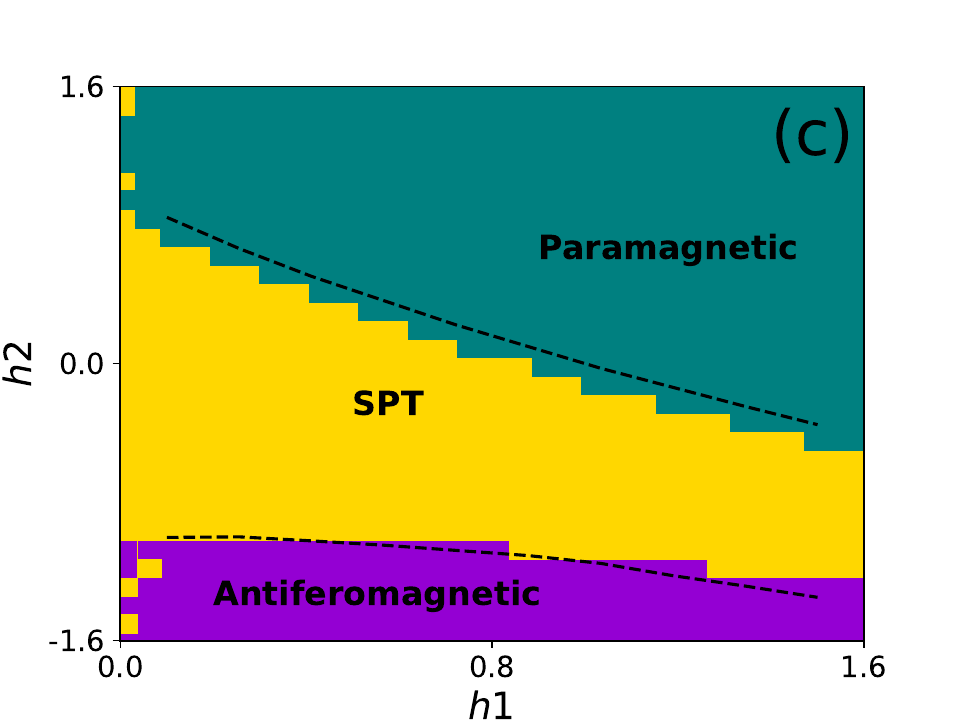} 
    \caption{ 
Results for the Cluster-Ising Hamiltonian \eqref{eq:Htopo} with $N = 51$ spins, where 
	the ground states are simulated with a bond dimension $\chi = 70$. 
    (a) Average silhouette score as a 
    function of the number of clusters $c$; the inset 
    shows the elbow point, identified by a significant change 
    in the WCSS dependence on $c$. (b) Silhouette plot for $c=3$, as determined by Fig.~\ref{fig:topo}a. 
  (c)~The predicted phase diagram.
    }
\label{fig:topo}
\end{figure*}

Fig.~\ref{fig:topo} presents our numerical results for a chain of $N = 51$ spins, sampling $30\times30$ pairs $(h_1,h_2)$.
We use the silhouette method to determine the number of 
phases,  i.e. the optimal number of clusters, $c$. 
The silohuette score as a function of $c$ is shown in Fig.~\ref{fig:topo}a, with the 
highest score at $c=3$. 
The WCSS vs.~$c$ shown in the inset confirms this result, according to the Elbow method.
The silhouette plot is shown in Fig.~\ref{fig:topo}b, where the dashed vertical line marks the 
average silhouette score. The 
results demonstrate that phase boundaries are identified even when the elusive SPT phase 
is involved, as silhouette scores are mostly positive.
However, some misclusterings are observed in the paramagnetic and antiferromagnetic phases, 
indicated by few negative silhouette scores.
Finally, the QPD for the model \eqref{eq:Htopo} is shown in Fig.~\ref{fig:topo}c.
Its structure closely matches the expected one, according to the above mentioned 
previous results~\cite{cong2019quantum, herrmann2022realizing}, represented by the black dashed lines. Minor discrepancies near the transition lines may arise from finite-size effects or limitations in the DMRG bond dimension used.

\section{Conclusion and Discussion\label{sec:conclusion}} 
We explored the use of classical unsupervised machine 
learning methods for studying QPD of quantum many-body systems. 
We found that kernel-based spectral clustering methods, where
the different clusters define the different quantum phases,
are effective when 
fed by kernels of genuine quantum origin.
Importantly, the number of phases is discovered by the algorithm itself. 

Two specific models of spin-$\frac{1}{2}$ chains, the ANNNI and 
the Cluster-Ising models, the latter of which featuring a topological
phase, are used to benchmark our results.
The bond dimension $\chi$ is a free parameter that defines the noise in the dataset, in the language of machine learning. 
For the ANNNI model, a moderate 
bond dimension ($\chi=20$) is sufficient, whereas the Cluster–Ising model requires a relatively larger 
bond dimension ($\chi=70$). 
It is relevant that the clustering approach 
successfully detects complex phases, such as those lacking a local order parameter,
demonstrating its ability to capture nontrivial topological features directly 
from quantum-generated datasets.

Our findings show that integrating classical unsupervised machine learning
models with quantum-generated datasets can provide physicists with a powerful tool 
for recognizing and classifying quantum phases of matter. 
Our work paves the way to the study of systems with completely unknown phase diagrams, without any prior 
labeling.  Amongst these, spin liquids~\cite{zhou2017quantum,wen2019experimental}, where conventional theoretical 
or numerical guidance is limited, and spin glasses~\cite{binder1986spin}, represent a promising avenue for future exploration, 
offering an opportunity to extend our methods to even more challenging quantum phases.

Another future direction is to replace our quantum kernels, estimated via tensor networks, with 
kernels directly reconstructed from experimental data, possibly using classical shadow techniques \cite{huang2020predicting}
or other randomized measurements \cite{cieslinski2024analysing}.

\section{Acknowledgment}
We thank Giuseppe Magnifico for valuable discussion.
All authors acknowledge financial support from: PNRR Ministero Universit\`a e Ricerca Project No. PE0000023-NQSTI funded by European Union-Next-Generation EU.
L.B.~also acknowledges financial support from: 
 Prin 2022 - DD N. 104 del 2/2/2022, entitled ``understanding the LEarning process of QUantum Neural networks (LeQun)'', proposal code 2022WHZ5XH, CUP B53D23009530006; 
 the European Union's Horizon Europe research and innovation program under EPIQUE Project GA No. 101135288. 
 This work is done in the framework of the Convenzione operativa between
the Institute for Complex Systems of the Consiglio Nazionale delle
Ricerche (Italy) and the Physics and Astronomy Department of the
University of Florence
\bibliography{biblio}

\begin{thebibliography}{70}%
\makeatletter
\providecommand \@ifxundefined [1]{%
 \@ifx{#1\undefined}
}%
\providecommand \@ifnum [1]{%
 \ifnum #1\expandafter \@firstoftwo
 \else \expandafter \@secondoftwo
 \fi
}%
\providecommand \@ifx [1]{%
 \ifx #1\expandafter \@firstoftwo
 \else \expandafter \@secondoftwo
 \fi
}%
\providecommand \natexlab [1]{#1}%
\providecommand \enquote  [1]{``#1''}%
\providecommand \bibnamefont  [1]{#1}%
\providecommand \bibfnamefont [1]{#1}%
\providecommand \citenamefont [1]{#1}%
\providecommand \href@noop [0]{\@secondoftwo}%
\providecommand \href [0]{\begingroup \@sanitize@url \@href}%
\providecommand \@href[1]{\@@startlink{#1}\@@href}%
\providecommand \@@href[1]{\endgroup#1\@@endlink}%
\providecommand \@sanitize@url [0]{\catcode `\\12\catcode `\$12\catcode `\&12\catcode `\#12\catcode `\^12\catcode `\_12\catcode `\%12\relax}%
\providecommand \@@startlink[1]{}%
\providecommand \@@endlink[0]{}%
\providecommand \url  [0]{\begingroup\@sanitize@url \@url }%
\providecommand \@url [1]{\endgroup\@href {#1}{\urlprefix }}%
\providecommand \urlprefix  [0]{URL }%
\providecommand \Eprint [0]{\href }%
\providecommand \doibase [0]{https://doi.org/}%
\providecommand \selectlanguage [0]{\@gobble}%
\providecommand \bibinfo  [0]{\@secondoftwo}%
\providecommand \bibfield  [0]{\@secondoftwo}%
\providecommand \translation [1]{[#1]}%
\providecommand \BibitemOpen [0]{}%
\providecommand \bibitemStop [0]{}%
\providecommand \bibitemNoStop [0]{.\EOS\space}%
\providecommand \EOS [0]{\spacefactor3000\relax}%
\providecommand \BibitemShut  [1]{\csname bibitem#1\endcsname}%
\let\auto@bib@innerbib\@empty
\bibitem [{\citenamefont {Sachdev}(1999)}]{sachdev1999quantum}%
  \BibitemOpen
  \bibfield  {author} {\bibinfo {author} {\bibfnamefont {S.}~\bibnamefont {Sachdev}},\ }\bibfield  {title} {\bibinfo {title} {Quantum phase transitions},\ }\href@noop {} {\bibfield  {journal} {\bibinfo  {journal} {Physics world}\ }\textbf {\bibinfo {volume} {12}},\ \bibinfo {pages} {33} (\bibinfo {year} {1999})}\BibitemShut {NoStop}%
\bibitem [{\citenamefont {Khosrojerdi}\ \emph {et~al.}(2025)\citenamefont {Khosrojerdi}, \citenamefont {Pereira}, \citenamefont {Cuccoli},\ and\ \citenamefont {Banchi}}]{Khosrojerdi_2025}%
  \BibitemOpen
  \bibfield  {author} {\bibinfo {author} {\bibfnamefont {M.}~\bibnamefont {Khosrojerdi}}, \bibinfo {author} {\bibfnamefont {J.~L.}\ \bibnamefont {Pereira}}, \bibinfo {author} {\bibfnamefont {A.}~\bibnamefont {Cuccoli}},\ and\ \bibinfo {author} {\bibfnamefont {L.}~\bibnamefont {Banchi}},\ }\bibfield  {title} {\bibinfo {title} {Learning to classify quantum phases of matter with a few measurements},\ }\href {https://doi.org/10.1088/2058-9565/ada79b} {\bibfield  {journal} {\bibinfo  {journal} {Quantum Science and Technology}\ }\textbf {\bibinfo {volume} {10}},\ \bibinfo {pages} {025006} (\bibinfo {year} {2025})}\BibitemShut {NoStop}%
\bibitem [{\citenamefont {Cong}\ \emph {et~al.}(2019)\citenamefont {Cong}, \citenamefont {Choi},\ and\ \citenamefont {Lukin}}]{cong2019quantum}%
  \BibitemOpen
  \bibfield  {author} {\bibinfo {author} {\bibfnamefont {I.}~\bibnamefont {Cong}}, \bibinfo {author} {\bibfnamefont {S.}~\bibnamefont {Choi}},\ and\ \bibinfo {author} {\bibfnamefont {M.~D.}\ \bibnamefont {Lukin}},\ }\bibfield  {title} {\bibinfo {title} {Quantum convolutional neural networks},\ }\href@noop {} {\bibfield  {journal} {\bibinfo  {journal} {Nature Physics}\ }\textbf {\bibinfo {volume} {15}},\ \bibinfo {pages} {1273} (\bibinfo {year} {2019})}\BibitemShut {NoStop}%
\bibitem [{\citenamefont {Monaco}\ \emph {et~al.}(2023)\citenamefont {Monaco}, \citenamefont {Kiss}, \citenamefont {Mandarino}, \citenamefont {Vallecorsa},\ and\ \citenamefont {Grossi}}]{monaco2023quantum}%
  \BibitemOpen
  \bibfield  {author} {\bibinfo {author} {\bibfnamefont {S.}~\bibnamefont {Monaco}}, \bibinfo {author} {\bibfnamefont {O.}~\bibnamefont {Kiss}}, \bibinfo {author} {\bibfnamefont {A.}~\bibnamefont {Mandarino}}, \bibinfo {author} {\bibfnamefont {S.}~\bibnamefont {Vallecorsa}},\ and\ \bibinfo {author} {\bibfnamefont {M.}~\bibnamefont {Grossi}},\ }\bibfield  {title} {\bibinfo {title} {Quantum phase detection generalization from marginal quantum neural network models},\ }\href@noop {} {\bibfield  {journal} {\bibinfo  {journal} {Physical Review B}\ }\textbf {\bibinfo {volume} {107}},\ \bibinfo {pages} {L081105} (\bibinfo {year} {2023})}\BibitemShut {NoStop}%
\bibitem [{\citenamefont {Huang}\ \emph {et~al.}(2022)\citenamefont {Huang}, \citenamefont {Kueng}, \citenamefont {Torlai}, \citenamefont {Albert},\ and\ \citenamefont {Preskill}}]{huang2022provably}%
  \BibitemOpen
  \bibfield  {author} {\bibinfo {author} {\bibfnamefont {H.-Y.}\ \bibnamefont {Huang}}, \bibinfo {author} {\bibfnamefont {R.}~\bibnamefont {Kueng}}, \bibinfo {author} {\bibfnamefont {G.}~\bibnamefont {Torlai}}, \bibinfo {author} {\bibfnamefont {V.~V.}\ \bibnamefont {Albert}},\ and\ \bibinfo {author} {\bibfnamefont {J.}~\bibnamefont {Preskill}},\ }\bibfield  {title} {\bibinfo {title} {Provably efficient machine learning for quantum many-body problems},\ }\href@noop {} {\bibfield  {journal} {\bibinfo  {journal} {Science}\ }\textbf {\bibinfo {volume} {377}},\ \bibinfo {pages} {eabk3333} (\bibinfo {year} {2022})}\BibitemShut {NoStop}%
\bibitem [{\citenamefont {Dong}\ \emph {et~al.}(2019)\citenamefont {Dong}, \citenamefont {Pollmann},\ and\ \citenamefont {Zhang}}]{dong2019machine}%
  \BibitemOpen
  \bibfield  {author} {\bibinfo {author} {\bibfnamefont {X.-Y.}\ \bibnamefont {Dong}}, \bibinfo {author} {\bibfnamefont {F.}~\bibnamefont {Pollmann}},\ and\ \bibinfo {author} {\bibfnamefont {X.-F.}\ \bibnamefont {Zhang}},\ }\bibfield  {title} {\bibinfo {title} {Machine learning of quantum phase transitions},\ }\href@noop {} {\bibfield  {journal} {\bibinfo  {journal} {Physical Review B}\ }\textbf {\bibinfo {volume} {99}},\ \bibinfo {pages} {121104} (\bibinfo {year} {2019})}\BibitemShut {NoStop}%
\bibitem [{\citenamefont {Carleo}\ \emph {et~al.}(2019)\citenamefont {Carleo}, \citenamefont {Cirac}, \citenamefont {Cranmer}, \citenamefont {Daudet}, \citenamefont {Schuld}, \citenamefont {Tishby}, \citenamefont {Vogt-Maranto},\ and\ \citenamefont {Zdeborov{\'a}}}]{carleo2019machine}%
  \BibitemOpen
  \bibfield  {author} {\bibinfo {author} {\bibfnamefont {G.}~\bibnamefont {Carleo}}, \bibinfo {author} {\bibfnamefont {I.}~\bibnamefont {Cirac}}, \bibinfo {author} {\bibfnamefont {K.}~\bibnamefont {Cranmer}}, \bibinfo {author} {\bibfnamefont {L.}~\bibnamefont {Daudet}}, \bibinfo {author} {\bibfnamefont {M.}~\bibnamefont {Schuld}}, \bibinfo {author} {\bibfnamefont {N.}~\bibnamefont {Tishby}}, \bibinfo {author} {\bibfnamefont {L.}~\bibnamefont {Vogt-Maranto}},\ and\ \bibinfo {author} {\bibfnamefont {L.}~\bibnamefont {Zdeborov{\'a}}},\ }\bibfield  {title} {\bibinfo {title} {Machine learning and the physical sciences},\ }\href@noop {} {\bibfield  {journal} {\bibinfo  {journal} {Reviews of Modern Physics}\ }\textbf {\bibinfo {volume} {91}},\ \bibinfo {pages} {045002} (\bibinfo {year} {2019})}\BibitemShut {NoStop}%
\bibitem [{\citenamefont {Cea}\ \emph {et~al.}(2024)\citenamefont {Cea}, \citenamefont {Grossi}, \citenamefont {Monaco}, \citenamefont {Rico}, \citenamefont {Tagliacozzo},\ and\ \citenamefont {Vallecorsa}}]{cea2024exploring}%
  \BibitemOpen
  \bibfield  {author} {\bibinfo {author} {\bibfnamefont {M.}~\bibnamefont {Cea}}, \bibinfo {author} {\bibfnamefont {M.}~\bibnamefont {Grossi}}, \bibinfo {author} {\bibfnamefont {S.}~\bibnamefont {Monaco}}, \bibinfo {author} {\bibfnamefont {E.}~\bibnamefont {Rico}}, \bibinfo {author} {\bibfnamefont {L.}~\bibnamefont {Tagliacozzo}},\ and\ \bibinfo {author} {\bibfnamefont {S.}~\bibnamefont {Vallecorsa}},\ }\bibfield  {title} {\bibinfo {title} {Exploring the phase diagram of the quantum one-dimensional annni model},\ }\href@noop {} {\bibfield  {journal} {\bibinfo  {journal} {arXiv preprint arXiv:2402.11022}\ } (\bibinfo {year} {2024})}\BibitemShut {NoStop}%
\bibitem [{\citenamefont {Uvarov}\ \emph {et~al.}(2020)\citenamefont {Uvarov}, \citenamefont {Kardashin},\ and\ \citenamefont {Biamonte}}]{uvarov2020machine}%
  \BibitemOpen
  \bibfield  {author} {\bibinfo {author} {\bibfnamefont {A.}~\bibnamefont {Uvarov}}, \bibinfo {author} {\bibfnamefont {A.}~\bibnamefont {Kardashin}},\ and\ \bibinfo {author} {\bibfnamefont {J.~D.}\ \bibnamefont {Biamonte}},\ }\bibfield  {title} {\bibinfo {title} {Machine learning phase transitions with a quantum processor},\ }\href@noop {} {\bibfield  {journal} {\bibinfo  {journal} {Physical Review A}\ }\textbf {\bibinfo {volume} {102}},\ \bibinfo {pages} {012415} (\bibinfo {year} {2020})}\BibitemShut {NoStop}%
\bibitem [{\citenamefont {Carrasquilla}\ and\ \citenamefont {Melko}(2017)}]{carrasquilla2017machine}%
  \BibitemOpen
  \bibfield  {author} {\bibinfo {author} {\bibfnamefont {J.}~\bibnamefont {Carrasquilla}}\ and\ \bibinfo {author} {\bibfnamefont {R.~G.}\ \bibnamefont {Melko}},\ }\bibfield  {title} {\bibinfo {title} {Machine learning phases of matter},\ }\href@noop {} {\bibfield  {journal} {\bibinfo  {journal} {Nature Physics}\ }\textbf {\bibinfo {volume} {13}},\ \bibinfo {pages} {431} (\bibinfo {year} {2017})}\BibitemShut {NoStop}%
\bibitem [{\citenamefont {Li}\ \emph {et~al.}(2024)\citenamefont {Li}, \citenamefont {Huang}, \citenamefont {Hou}, \citenamefont {Li}, \citenamefont {Wang},\ and\ \citenamefont {Bayat}}]{li2024ensemble}%
  \BibitemOpen
  \bibfield  {author} {\bibinfo {author} {\bibfnamefont {Q.}~\bibnamefont {Li}}, \bibinfo {author} {\bibfnamefont {Y.}~\bibnamefont {Huang}}, \bibinfo {author} {\bibfnamefont {X.}~\bibnamefont {Hou}}, \bibinfo {author} {\bibfnamefont {Y.}~\bibnamefont {Li}}, \bibinfo {author} {\bibfnamefont {X.}~\bibnamefont {Wang}},\ and\ \bibinfo {author} {\bibfnamefont {A.}~\bibnamefont {Bayat}},\ }\bibfield  {title} {\bibinfo {title} {Ensemble-learning error mitigation for variational quantum shallow-circuit classifiers},\ }\href@noop {} {\bibfield  {journal} {\bibinfo  {journal} {Physical Review Research}\ }\textbf {\bibinfo {volume} {6}},\ \bibinfo {pages} {013027} (\bibinfo {year} {2024})}\BibitemShut {NoStop}%
\bibitem [{\citenamefont {Parigi}\ \emph {et~al.}(2025)\citenamefont {Parigi}, \citenamefont {Khosrojerdi}, \citenamefont {Caruso},\ and\ \citenamefont {Banchi}}]{parigi2025supervised}%
  \BibitemOpen
  \bibfield  {author} {\bibinfo {author} {\bibfnamefont {M.}~\bibnamefont {Parigi}}, \bibinfo {author} {\bibfnamefont {M.}~\bibnamefont {Khosrojerdi}}, \bibinfo {author} {\bibfnamefont {F.}~\bibnamefont {Caruso}},\ and\ \bibinfo {author} {\bibfnamefont {L.}~\bibnamefont {Banchi}},\ }\bibfield  {title} {\bibinfo {title} {Supervised quantum image processing},\ }\href@noop {} {\bibfield  {journal} {\bibinfo  {journal} {arXiv preprint arXiv:2507.22039}\ } (\bibinfo {year} {2025})}\BibitemShut {NoStop}%
\bibitem [{\citenamefont {Sent{\'\i}s}\ \emph {et~al.}(2019)\citenamefont {Sent{\'\i}s}, \citenamefont {Monras}, \citenamefont {Mu{\~n}oz-Tapia}, \citenamefont {Calsamiglia},\ and\ \citenamefont {Bagan}}]{sentis2019unsupervised}%
  \BibitemOpen
  \bibfield  {author} {\bibinfo {author} {\bibfnamefont {G.}~\bibnamefont {Sent{\'\i}s}}, \bibinfo {author} {\bibfnamefont {A.}~\bibnamefont {Monras}}, \bibinfo {author} {\bibfnamefont {R.}~\bibnamefont {Mu{\~n}oz-Tapia}}, \bibinfo {author} {\bibfnamefont {J.}~\bibnamefont {Calsamiglia}},\ and\ \bibinfo {author} {\bibfnamefont {E.}~\bibnamefont {Bagan}},\ }\bibfield  {title} {\bibinfo {title} {Unsupervised classification of quantum data},\ }\href@noop {} {\bibfield  {journal} {\bibinfo  {journal} {Physical Review X}\ }\textbf {\bibinfo {volume} {9}},\ \bibinfo {pages} {041029} (\bibinfo {year} {2019})}\BibitemShut {NoStop}%
\bibitem [{\citenamefont {Chen}\ \emph {et~al.}(2021)\citenamefont {Chen}, \citenamefont {Pan}, \citenamefont {Zhang},\ and\ \citenamefont {Cheng}}]{chen2021detecting}%
  \BibitemOpen
  \bibfield  {author} {\bibinfo {author} {\bibfnamefont {Y.}~\bibnamefont {Chen}}, \bibinfo {author} {\bibfnamefont {Y.}~\bibnamefont {Pan}}, \bibinfo {author} {\bibfnamefont {G.}~\bibnamefont {Zhang}},\ and\ \bibinfo {author} {\bibfnamefont {S.}~\bibnamefont {Cheng}},\ }\bibfield  {title} {\bibinfo {title} {Detecting quantum entanglement with unsupervised learning},\ }\href@noop {} {\bibfield  {journal} {\bibinfo  {journal} {Quantum Science and Technology}\ }\textbf {\bibinfo {volume} {7}},\ \bibinfo {pages} {015005} (\bibinfo {year} {2021})}\BibitemShut {NoStop}%
\bibitem [{\citenamefont {Yang}\ \emph {et~al.}(2021)\citenamefont {Yang}, \citenamefont {Sun}, \citenamefont {Ran},\ and\ \citenamefont {Su}}]{yang2021visualizing}%
  \BibitemOpen
  \bibfield  {author} {\bibinfo {author} {\bibfnamefont {Y.}~\bibnamefont {Yang}}, \bibinfo {author} {\bibfnamefont {Z.-Z.}\ \bibnamefont {Sun}}, \bibinfo {author} {\bibfnamefont {S.-J.}\ \bibnamefont {Ran}},\ and\ \bibinfo {author} {\bibfnamefont {G.}~\bibnamefont {Su}},\ }\bibfield  {title} {\bibinfo {title} {Visualizing quantum phases and identifying quantum phase transitions by nonlinear dimensional reduction},\ }\href@noop {} {\bibfield  {journal} {\bibinfo  {journal} {Physical Review B}\ }\textbf {\bibinfo {volume} {103}},\ \bibinfo {pages} {075106} (\bibinfo {year} {2021})}\BibitemShut {NoStop}%
\bibitem [{\citenamefont {Tibaldi}\ \emph {et~al.}(2023)\citenamefont {Tibaldi}, \citenamefont {Magnifico}, \citenamefont {Vodola},\ and\ \citenamefont {Ercolessi}}]{10.21468/SciPostPhys.14.1.005}%
  \BibitemOpen
  \bibfield  {author} {\bibinfo {author} {\bibfnamefont {S.}~\bibnamefont {Tibaldi}}, \bibinfo {author} {\bibfnamefont {G.}~\bibnamefont {Magnifico}}, \bibinfo {author} {\bibfnamefont {D.}~\bibnamefont {Vodola}},\ and\ \bibinfo {author} {\bibfnamefont {E.}~\bibnamefont {Ercolessi}},\ }\bibfield  {title} {\bibinfo {title} {{Unsupervised and supervised learning of interacting topological phases from single-particle correlation functions}},\ }\href {https://doi.org/10.21468/SciPostPhys.14.1.005} {\bibfield  {journal} {\bibinfo  {journal} {SciPost Phys.}\ }\textbf {\bibinfo {volume} {14}},\ \bibinfo {pages} {005} (\bibinfo {year} {2023})}\BibitemShut {NoStop}%
\bibitem [{\citenamefont {Zinn-Justin}(2007)}]{zinn2007phase}%
  \BibitemOpen
  \bibfield  {author} {\bibinfo {author} {\bibfnamefont {J.}~\bibnamefont {Zinn-Justin}},\ }\href@noop {} {\emph {\bibinfo {title} {Phase transitions and renormalization group}}}\ (\bibinfo  {publisher} {Oxford University Press},\ \bibinfo {year} {2007})\BibitemShut {NoStop}%
\bibitem [{\citenamefont {Ng}\ \emph {et~al.}(2001)\citenamefont {Ng}, \citenamefont {Jordan},\ and\ \citenamefont {Weiss}}]{ng2001spectral}%
  \BibitemOpen
  \bibfield  {author} {\bibinfo {author} {\bibfnamefont {A.}~\bibnamefont {Ng}}, \bibinfo {author} {\bibfnamefont {M.}~\bibnamefont {Jordan}},\ and\ \bibinfo {author} {\bibfnamefont {Y.}~\bibnamefont {Weiss}},\ }\bibfield  {title} {\bibinfo {title} {On spectral clustering: Analysis and an algorithm},\ }\href@noop {} {\bibfield  {journal} {\bibinfo  {journal} {Advances in neural information processing systems}\ }\textbf {\bibinfo {volume} {14}} (\bibinfo {year} {2001})}\BibitemShut {NoStop}%
\bibitem [{\citenamefont {Von~Luxburg}(2007)}]{von2007tutorial}%
  \BibitemOpen
  \bibfield  {author} {\bibinfo {author} {\bibfnamefont {U.}~\bibnamefont {Von~Luxburg}},\ }\bibfield  {title} {\bibinfo {title} {A tutorial on spectral clustering},\ }\href@noop {} {\bibfield  {journal} {\bibinfo  {journal} {Statistics and computing}\ }\textbf {\bibinfo {volume} {17}},\ \bibinfo {pages} {395} (\bibinfo {year} {2007})}\BibitemShut {NoStop}%
\bibitem [{\citenamefont {Zhou}\ and\ \citenamefont {A.Amini}(2019)}]{JMLR:v20:18-170}%
  \BibitemOpen
  \bibfield  {author} {\bibinfo {author} {\bibfnamefont {Z.}~\bibnamefont {Zhou}}\ and\ \bibinfo {author} {\bibfnamefont {A.}~\bibnamefont {A.Amini}},\ }\bibfield  {title} {\bibinfo {title} {Analysis of spectral clustering algorithms for community detection: the general bipartite setting},\ }\href {http://jmlr.org/papers/v20/18-170.html} {\bibfield  {journal} {\bibinfo  {journal} {Journal of Machine Learning Research}\ }\textbf {\bibinfo {volume} {20}},\ \bibinfo {pages} {1} (\bibinfo {year} {2019})}\BibitemShut {NoStop}%
\bibitem [{\citenamefont {Schiebinger}\ \emph {et~al.}(2015)\citenamefont {Schiebinger}, \citenamefont {Wainwright},\ and\ \citenamefont {Yu}}]{10.1214/14-AOS1283}%
  \BibitemOpen
  \bibfield  {author} {\bibinfo {author} {\bibfnamefont {G.}~\bibnamefont {Schiebinger}}, \bibinfo {author} {\bibfnamefont {M.~J.}\ \bibnamefont {Wainwright}},\ and\ \bibinfo {author} {\bibfnamefont {B.}~\bibnamefont {Yu}},\ }\bibfield  {title} {\bibinfo {title} {{The geometry of kernelized spectral clustering}},\ }\href {https://doi.org/10.1214/14-AOS1283} {\bibfield  {journal} {\bibinfo  {journal} {The Annals of Statistics}\ }\textbf {\bibinfo {volume} {43}},\ \bibinfo {pages} {819 } (\bibinfo {year} {2015})}\BibitemShut {NoStop}%
\bibitem [{\citenamefont {Bengio}\ \emph {et~al.}(2004)\citenamefont {Bengio}, \citenamefont {Delalleau}, \citenamefont {Roux}, \citenamefont {Paiement}, \citenamefont {Vincent},\ and\ \citenamefont {Ouimet}}]{10.1162/0899766041732396}%
  \BibitemOpen
  \bibfield  {author} {\bibinfo {author} {\bibfnamefont {Y.}~\bibnamefont {Bengio}}, \bibinfo {author} {\bibfnamefont {O.}~\bibnamefont {Delalleau}}, \bibinfo {author} {\bibfnamefont {N.~L.}\ \bibnamefont {Roux}}, \bibinfo {author} {\bibfnamefont {J.-F.}\ \bibnamefont {Paiement}}, \bibinfo {author} {\bibfnamefont {P.}~\bibnamefont {Vincent}},\ and\ \bibinfo {author} {\bibfnamefont {M.}~\bibnamefont {Ouimet}},\ }\bibfield  {title} {\bibinfo {title} {Learning eigenfunctions links spectral embedding and kernel pca},\ }\href@noop {} {\bibfield  {journal} {\bibinfo  {journal} {Neural Computation}\ }\textbf {\bibinfo {volume} {16}},\ \bibinfo {pages} {2197} (\bibinfo {year} {2004})}\BibitemShut {NoStop}%
\bibitem [{\citenamefont {Perez-Garcia}\ \emph {et~al.}(2006)\citenamefont {Perez-Garcia}, \citenamefont {Verstraete}, \citenamefont {Wolf},\ and\ \citenamefont {Cirac}}]{perez2006matrix}%
  \BibitemOpen
  \bibfield  {author} {\bibinfo {author} {\bibfnamefont {D.}~\bibnamefont {Perez-Garcia}}, \bibinfo {author} {\bibfnamefont {F.}~\bibnamefont {Verstraete}}, \bibinfo {author} {\bibfnamefont {M.~M.}\ \bibnamefont {Wolf}},\ and\ \bibinfo {author} {\bibfnamefont {J.~I.}\ \bibnamefont {Cirac}},\ }\bibfield  {title} {\bibinfo {title} {Matrix product state representations},\ }\href@noop {} {\bibfield  {journal} {\bibinfo  {journal} {arXiv preprint quant-ph/0608197}\ } (\bibinfo {year} {2006})}\BibitemShut {NoStop}%
\bibitem [{\citenamefont {Or{\'u}s}(2014)}]{orus2014practical}%
  \BibitemOpen
  \bibfield  {author} {\bibinfo {author} {\bibfnamefont {R.}~\bibnamefont {Or{\'u}s}},\ }\bibfield  {title} {\bibinfo {title} {A practical introduction to tensor networks: Matrix product states and projected entangled pair states},\ }\href@noop {} {\bibfield  {journal} {\bibinfo  {journal} {Annals of physics}\ }\textbf {\bibinfo {volume} {349}},\ \bibinfo {pages} {117} (\bibinfo {year} {2014})}\BibitemShut {NoStop}%
\bibitem [{\citenamefont {Ran}\ \emph {et~al.}(2020)\citenamefont {Ran}, \citenamefont {Tirrito}, \citenamefont {Peng}, \citenamefont {Chen}, \citenamefont {Tagliacozzo}, \citenamefont {Su},\ and\ \citenamefont {Lewenstein}}]{ran2020tensor}%
  \BibitemOpen
  \bibfield  {author} {\bibinfo {author} {\bibfnamefont {S.-J.}\ \bibnamefont {Ran}}, \bibinfo {author} {\bibfnamefont {E.}~\bibnamefont {Tirrito}}, \bibinfo {author} {\bibfnamefont {C.}~\bibnamefont {Peng}}, \bibinfo {author} {\bibfnamefont {X.}~\bibnamefont {Chen}}, \bibinfo {author} {\bibfnamefont {L.}~\bibnamefont {Tagliacozzo}}, \bibinfo {author} {\bibfnamefont {G.}~\bibnamefont {Su}},\ and\ \bibinfo {author} {\bibfnamefont {M.}~\bibnamefont {Lewenstein}},\ }\href@noop {} {\emph {\bibinfo {title} {Tensor network contractions: methods and applications to quantum many-body systems}}}\ (\bibinfo  {publisher} {Springer Nature},\ \bibinfo {year} {2020})\BibitemShut {NoStop}%
\bibitem [{\citenamefont {Schuld}\ and\ \citenamefont {Petruccione}(2021)}]{schuld2021machine}%
  \BibitemOpen
  \bibfield  {author} {\bibinfo {author} {\bibfnamefont {M.}~\bibnamefont {Schuld}}\ and\ \bibinfo {author} {\bibfnamefont {F.}~\bibnamefont {Petruccione}},\ }\href@noop {} {\emph {\bibinfo {title} {Machine learning with quantum computers}}},\ Vol.\ \bibinfo {volume} {676}\ (\bibinfo  {publisher} {Springer},\ \bibinfo {year} {2021})\BibitemShut {NoStop}%
\bibitem [{\citenamefont {Tilly}\ \emph {et~al.}(2022)\citenamefont {Tilly}, \citenamefont {Chen}, \citenamefont {Cao}, \citenamefont {Picozzi}, \citenamefont {Setia}, \citenamefont {Li}, \citenamefont {Grant}, \citenamefont {Wossnig}, \citenamefont {Rungger}, \citenamefont {Booth} \emph {et~al.}}]{tilly2022variational}%
  \BibitemOpen
  \bibfield  {author} {\bibinfo {author} {\bibfnamefont {J.}~\bibnamefont {Tilly}}, \bibinfo {author} {\bibfnamefont {H.}~\bibnamefont {Chen}}, \bibinfo {author} {\bibfnamefont {S.}~\bibnamefont {Cao}}, \bibinfo {author} {\bibfnamefont {D.}~\bibnamefont {Picozzi}}, \bibinfo {author} {\bibfnamefont {K.}~\bibnamefont {Setia}}, \bibinfo {author} {\bibfnamefont {Y.}~\bibnamefont {Li}}, \bibinfo {author} {\bibfnamefont {E.}~\bibnamefont {Grant}}, \bibinfo {author} {\bibfnamefont {L.}~\bibnamefont {Wossnig}}, \bibinfo {author} {\bibfnamefont {I.}~\bibnamefont {Rungger}}, \bibinfo {author} {\bibfnamefont {G.~H.}\ \bibnamefont {Booth}}, \emph {et~al.},\ }\bibfield  {title} {\bibinfo {title} {The variational quantum eigensolver: a review of methods and best practices},\ }\href@noop {} {\bibfield  {journal} {\bibinfo  {journal} {Physics Reports}\ }\textbf {\bibinfo {volume} {986}},\ \bibinfo {pages} {1} (\bibinfo {year} {2022})}\BibitemShut {NoStop}%
\bibitem [{\citenamefont {Grimsley}\ \emph {et~al.}(2019)\citenamefont {Grimsley}, \citenamefont {Economou}, \citenamefont {Barnes},\ and\ \citenamefont {Mayhall}}]{grimsley2019adaptive}%
  \BibitemOpen
  \bibfield  {author} {\bibinfo {author} {\bibfnamefont {H.~R.}\ \bibnamefont {Grimsley}}, \bibinfo {author} {\bibfnamefont {S.~E.}\ \bibnamefont {Economou}}, \bibinfo {author} {\bibfnamefont {E.}~\bibnamefont {Barnes}},\ and\ \bibinfo {author} {\bibfnamefont {N.~J.}\ \bibnamefont {Mayhall}},\ }\bibfield  {title} {\bibinfo {title} {An adaptive variational algorithm for exact molecular simulations on a quantum computer},\ }\href@noop {} {\bibfield  {journal} {\bibinfo  {journal} {Nature communications}\ }\textbf {\bibinfo {volume} {10}},\ \bibinfo {pages} {3007} (\bibinfo {year} {2019})}\BibitemShut {NoStop}%
\bibitem [{\citenamefont {Schuld}(2021)}]{schuld2021supervised}%
  \BibitemOpen
  \bibfield  {author} {\bibinfo {author} {\bibfnamefont {M.}~\bibnamefont {Schuld}},\ }\bibfield  {title} {\bibinfo {title} {Supervised quantum machine learning models are kernel methods},\ }\href@noop {} {\bibfield  {journal} {\bibinfo  {journal} {arXiv preprint arXiv:2101.11020}\ } (\bibinfo {year} {2021})}\BibitemShut {NoStop}%
\bibitem [{\citenamefont {Arizmendi}\ \emph {et~al.}(1991)\citenamefont {Arizmendi}, \citenamefont {Rizzo}, \citenamefont {Epele},\ and\ \citenamefont {Garc{\'\i}a~Canal}}]{arizmendi1991phase}%
  \BibitemOpen
  \bibfield  {author} {\bibinfo {author} {\bibfnamefont {C.}~\bibnamefont {Arizmendi}}, \bibinfo {author} {\bibfnamefont {A.}~\bibnamefont {Rizzo}}, \bibinfo {author} {\bibfnamefont {L.~N.}\ \bibnamefont {Epele}},\ and\ \bibinfo {author} {\bibfnamefont {C.~A.}\ \bibnamefont {Garc{\'\i}a~Canal}},\ }\bibfield  {title} {\bibinfo {title} {Phase diagram of the annni model in the hamiltonian limit},\ }\href@noop {} {\bibfield  {journal} {\bibinfo  {journal} {Zeitschrift f{\"u}r Physik B Condensed Matter}\ }\textbf {\bibinfo {volume} {83}},\ \bibinfo {pages} {273} (\bibinfo {year} {1991})}\BibitemShut {NoStop}%
\bibitem [{\citenamefont {Beccaria}\ \emph {et~al.}(2007{\natexlab{a}})\citenamefont {Beccaria}, \citenamefont {Campostrini},\ and\ \citenamefont {Feo}}]{PhysRevB.76.094410}%
  \BibitemOpen
  \bibfield  {author} {\bibinfo {author} {\bibfnamefont {M.}~\bibnamefont {Beccaria}}, \bibinfo {author} {\bibfnamefont {M.}~\bibnamefont {Campostrini}},\ and\ \bibinfo {author} {\bibfnamefont {A.}~\bibnamefont {Feo}},\ }\bibfield  {title} {\bibinfo {title} {Evidence for a floating phase of the transverse annni model at high frustration},\ }\href {https://doi.org/10.1103/PhysRevB.76.094410} {\bibfield  {journal} {\bibinfo  {journal} {Phys. Rev. B}\ }\textbf {\bibinfo {volume} {76}},\ \bibinfo {pages} {094410} (\bibinfo {year} {2007}{\natexlab{a}})}\BibitemShut {NoStop}%
\bibitem [{\citenamefont {Chandra}\ and\ \citenamefont {Dasgupta}(2007{\natexlab{a}})}]{PhysRevE.75.021105}%
  \BibitemOpen
  \bibfield  {author} {\bibinfo {author} {\bibfnamefont {A.~K.}\ \bibnamefont {Chandra}}\ and\ \bibinfo {author} {\bibfnamefont {S.}~\bibnamefont {Dasgupta}},\ }\bibfield  {title} {\bibinfo {title} {Floating phase in the one-dimensional transverse axial next-nearest-neighbor ising model},\ }\href {https://doi.org/10.1103/PhysRevE.75.021105} {\bibfield  {journal} {\bibinfo  {journal} {Phys. Rev. E}\ }\textbf {\bibinfo {volume} {75}},\ \bibinfo {pages} {021105} (\bibinfo {year} {2007}{\natexlab{a}})}\BibitemShut {NoStop}%
\bibitem [{\citenamefont {Nagy}(2011{\natexlab{a}})}]{Nagy_2011}%
  \BibitemOpen
  \bibfield  {author} {\bibinfo {author} {\bibfnamefont {A.}~\bibnamefont {Nagy}},\ }\bibfield  {title} {\bibinfo {title} {Exploring phase transitions by finite-entanglement scaling of mps in the 1d annni model},\ }\href {https://doi.org/10.1088/1367-2630/13/2/023015} {\bibfield  {journal} {\bibinfo  {journal} {New Journal of Physics}\ }\textbf {\bibinfo {volume} {13}},\ \bibinfo {pages} {023015} (\bibinfo {year} {2011}{\natexlab{a}})}\BibitemShut {NoStop}%
\bibitem [{\citenamefont {Caro}\ \emph {et~al.}(2023)\citenamefont {Caro}, \citenamefont {Huang}, \citenamefont {Ezzell}, \citenamefont {Gibbs}, \citenamefont {Sornborger}, \citenamefont {Cincio}, \citenamefont {Coles},\ and\ \citenamefont {Holmes}}]{caro2023out}%
  \BibitemOpen
  \bibfield  {author} {\bibinfo {author} {\bibfnamefont {M.~C.}\ \bibnamefont {Caro}}, \bibinfo {author} {\bibfnamefont {H.-Y.}\ \bibnamefont {Huang}}, \bibinfo {author} {\bibfnamefont {N.}~\bibnamefont {Ezzell}}, \bibinfo {author} {\bibfnamefont {J.}~\bibnamefont {Gibbs}}, \bibinfo {author} {\bibfnamefont {A.~T.}\ \bibnamefont {Sornborger}}, \bibinfo {author} {\bibfnamefont {L.}~\bibnamefont {Cincio}}, \bibinfo {author} {\bibfnamefont {P.~J.}\ \bibnamefont {Coles}},\ and\ \bibinfo {author} {\bibfnamefont {Z.}~\bibnamefont {Holmes}},\ }\bibfield  {title} {\bibinfo {title} {Out-of-distribution generalization for learning quantum dynamics},\ }\href@noop {} {\bibfield  {journal} {\bibinfo  {journal} {Nature Communications}\ }\textbf {\bibinfo {volume} {14}},\ \bibinfo {pages} {3751} (\bibinfo {year} {2023})}\BibitemShut {NoStop}%
\bibitem [{\citenamefont {Barouch}\ \emph {et~al.}(1970)\citenamefont {Barouch}, \citenamefont {McCoy},\ and\ \citenamefont {Dresden}}]{PhysRevA.2.1075}%
  \BibitemOpen
  \bibfield  {author} {\bibinfo {author} {\bibfnamefont {E.}~\bibnamefont {Barouch}}, \bibinfo {author} {\bibfnamefont {B.~M.}\ \bibnamefont {McCoy}},\ and\ \bibinfo {author} {\bibfnamefont {M.}~\bibnamefont {Dresden}},\ }\bibfield  {title} {\bibinfo {title} {Statistical mechanics of the $\mathrm{XY}$ model. i},\ }\href {https://doi.org/10.1103/PhysRevA.2.1075} {\bibfield  {journal} {\bibinfo  {journal} {Phys. Rev. A}\ }\textbf {\bibinfo {volume} {2}},\ \bibinfo {pages} {1075} (\bibinfo {year} {1970})}\BibitemShut {NoStop}%
\bibitem [{\citenamefont {Mbeng}\ \emph {et~al.}(2024)\citenamefont {Mbeng}, \citenamefont {Russomanno},\ and\ \citenamefont {Santoro}}]{mbeng2024quantum}%
  \BibitemOpen
  \bibfield  {author} {\bibinfo {author} {\bibfnamefont {G.~B.}\ \bibnamefont {Mbeng}}, \bibinfo {author} {\bibfnamefont {A.}~\bibnamefont {Russomanno}},\ and\ \bibinfo {author} {\bibfnamefont {G.~E.}\ \bibnamefont {Santoro}},\ }\bibfield  {title} {\bibinfo {title} {The quantum ising chain for beginners},\ }\href@noop {} {\bibfield  {journal} {\bibinfo  {journal} {SciPost Physics Lecture Notes}\ ,\ \bibinfo {pages} {082}} (\bibinfo {year} {2024})}\BibitemShut {NoStop}%
\bibitem [{\citenamefont {Parker}\ \emph {et~al.}(2020)\citenamefont {Parker}, \citenamefont {Cao},\ and\ \citenamefont {Zaletel}}]{parker2020local}%
  \BibitemOpen
  \bibfield  {author} {\bibinfo {author} {\bibfnamefont {D.~E.}\ \bibnamefont {Parker}}, \bibinfo {author} {\bibfnamefont {X.}~\bibnamefont {Cao}},\ and\ \bibinfo {author} {\bibfnamefont {M.~P.}\ \bibnamefont {Zaletel}},\ }\bibfield  {title} {\bibinfo {title} {Local matrix product operators: Canonical form, compression, and control theory},\ }\href@noop {} {\bibfield  {journal} {\bibinfo  {journal} {Physical Review B}\ }\textbf {\bibinfo {volume} {102}},\ \bibinfo {pages} {035147} (\bibinfo {year} {2020})}\BibitemShut {NoStop}%
\bibitem [{\citenamefont {Schollw{\"o}ck}(2011)}]{schollwock2011density}%
  \BibitemOpen
  \bibfield  {author} {\bibinfo {author} {\bibfnamefont {U.}~\bibnamefont {Schollw{\"o}ck}},\ }\bibfield  {title} {\bibinfo {title} {The density-matrix renormalization group in the age of matrix product states},\ }\href@noop {} {\bibfield  {journal} {\bibinfo  {journal} {Annals of physics}\ }\textbf {\bibinfo {volume} {326}},\ \bibinfo {pages} {96} (\bibinfo {year} {2011})}\BibitemShut {NoStop}%
\bibitem [{\citenamefont {Ikotun}\ \emph {et~al.}(2023)\citenamefont {Ikotun}, \citenamefont {Ezugwu}, \citenamefont {Abualigah}, \citenamefont {Abuhaija},\ and\ \citenamefont {Heming}}]{ikotun2023k}%
  \BibitemOpen
  \bibfield  {author} {\bibinfo {author} {\bibfnamefont {A.~M.}\ \bibnamefont {Ikotun}}, \bibinfo {author} {\bibfnamefont {A.~E.}\ \bibnamefont {Ezugwu}}, \bibinfo {author} {\bibfnamefont {L.}~\bibnamefont {Abualigah}}, \bibinfo {author} {\bibfnamefont {B.}~\bibnamefont {Abuhaija}},\ and\ \bibinfo {author} {\bibfnamefont {J.}~\bibnamefont {Heming}},\ }\bibfield  {title} {\bibinfo {title} {K-means clustering algorithms: A comprehensive review, variants analysis, and advances in the era of big data},\ }\href@noop {} {\bibfield  {journal} {\bibinfo  {journal} {Information Sciences}\ }\textbf {\bibinfo {volume} {622}},\ \bibinfo {pages} {178} (\bibinfo {year} {2023})}\BibitemShut {NoStop}%
\bibitem [{\citenamefont {Shi}\ \emph {et~al.}(2021)\citenamefont {Shi}, \citenamefont {Wei}, \citenamefont {Wei}, \citenamefont {Wang}, \citenamefont {Liu},\ and\ \citenamefont {Liu}}]{shi2021quantitative}%
  \BibitemOpen
  \bibfield  {author} {\bibinfo {author} {\bibfnamefont {C.}~\bibnamefont {Shi}}, \bibinfo {author} {\bibfnamefont {B.}~\bibnamefont {Wei}}, \bibinfo {author} {\bibfnamefont {S.}~\bibnamefont {Wei}}, \bibinfo {author} {\bibfnamefont {W.}~\bibnamefont {Wang}}, \bibinfo {author} {\bibfnamefont {H.}~\bibnamefont {Liu}},\ and\ \bibinfo {author} {\bibfnamefont {J.}~\bibnamefont {Liu}},\ }\bibfield  {title} {\bibinfo {title} {A quantitative discriminant method of elbow point for the optimal number of clusters in clustering algorithm},\ }\href@noop {} {\bibfield  {journal} {\bibinfo  {journal} {EURASIP journal on wireless communications and networking}\ }\textbf {\bibinfo {volume} {2021}},\ \bibinfo {pages} {31} (\bibinfo {year} {2021})}\BibitemShut {NoStop}%
\bibitem [{\citenamefont {Onumanyi}\ \emph {et~al.}(2022)\citenamefont {Onumanyi}, \citenamefont {Molokomme}, \citenamefont {Isaac},\ and\ \citenamefont {Abu-Mahfouz}}]{onumanyi2022autoelbow}%
  \BibitemOpen
  \bibfield  {author} {\bibinfo {author} {\bibfnamefont {A.~J.}\ \bibnamefont {Onumanyi}}, \bibinfo {author} {\bibfnamefont {D.~N.}\ \bibnamefont {Molokomme}}, \bibinfo {author} {\bibfnamefont {S.~J.}\ \bibnamefont {Isaac}},\ and\ \bibinfo {author} {\bibfnamefont {A.~M.}\ \bibnamefont {Abu-Mahfouz}},\ }\bibfield  {title} {\bibinfo {title} {Autoelbow: An automatic elbow detection method for estimating the number of clusters in a dataset},\ }\href@noop {} {\bibfield  {journal} {\bibinfo  {journal} {Applied Sciences}\ }\textbf {\bibinfo {volume} {12}},\ \bibinfo {pages} {7515} (\bibinfo {year} {2022})}\BibitemShut {NoStop}%
\bibitem [{\citenamefont {Rousseeuw}(1987{\natexlab{a}})}]{rousseeuw1987silhouettes}%
  \BibitemOpen
  \bibfield  {author} {\bibinfo {author} {\bibfnamefont {P.~J.}\ \bibnamefont {Rousseeuw}},\ }\bibfield  {title} {\bibinfo {title} {Silhouettes: a graphical aid to the interpretation and validation of cluster analysis},\ }\href@noop {} {\bibfield  {journal} {\bibinfo  {journal} {Journal of computational and applied mathematics}\ }\textbf {\bibinfo {volume} {20}},\ \bibinfo {pages} {53} (\bibinfo {year} {1987}{\natexlab{a}})}\BibitemShut {NoStop}%
\bibitem [{\citenamefont {Bholowalia}\ and\ \citenamefont {Kumar}(2014)}]{bholowalia2014ebk}%
  \BibitemOpen
  \bibfield  {author} {\bibinfo {author} {\bibfnamefont {P.}~\bibnamefont {Bholowalia}}\ and\ \bibinfo {author} {\bibfnamefont {A.}~\bibnamefont {Kumar}},\ }\bibfield  {title} {\bibinfo {title} {Ebk-means: A clustering technique based on elbow method and k-means in wsn},\ }\href@noop {} {\bibfield  {journal} {\bibinfo  {journal} {International Journal of Computer Applications}\ }\textbf {\bibinfo {volume} {105}} (\bibinfo {year} {2014})}\BibitemShut {NoStop}%
\bibitem [{\citenamefont {Kodinariya}\ \emph {et~al.}(2013)\citenamefont {Kodinariya}, \citenamefont {Makwana} \emph {et~al.}}]{kodinariya2013review}%
  \BibitemOpen
  \bibfield  {author} {\bibinfo {author} {\bibfnamefont {T.~M.}\ \bibnamefont {Kodinariya}}, \bibinfo {author} {\bibfnamefont {P.~R.}\ \bibnamefont {Makwana}}, \emph {et~al.},\ }\bibfield  {title} {\bibinfo {title} {Review on determining number of cluster in k-means clustering},\ }\href@noop {} {\bibfield  {journal} {\bibinfo  {journal} {International Journal}\ }\textbf {\bibinfo {volume} {1}},\ \bibinfo {pages} {90} (\bibinfo {year} {2013})}\BibitemShut {NoStop}%
\bibitem [{\citenamefont {Rousseeuw}(1987{\natexlab{b}})}]{ROUSSEEUW198753}%
  \BibitemOpen
  \bibfield  {author} {\bibinfo {author} {\bibfnamefont {P.~J.}\ \bibnamefont {Rousseeuw}},\ }\bibfield  {title} {\bibinfo {title} {Silhouettes: A graphical aid to the interpretation and validation of cluster analysis},\ }\href {https://doi.org/https://doi.org/10.1016/0377-0427(87)90125-7} {\bibfield  {journal} {\bibinfo  {journal} {Journal of Computational and Applied Mathematics}\ }\textbf {\bibinfo {volume} {20}},\ \bibinfo {pages} {53} (\bibinfo {year} {1987}{\natexlab{b}})}\BibitemShut {NoStop}%
\bibitem [{\citenamefont {Monshizadeh}\ \emph {et~al.}(2022)\citenamefont {Monshizadeh}, \citenamefont {Khatri}, \citenamefont {Kantola},\ and\ \citenamefont {Yan}}]{MONSHIZADEH2022103513}%
  \BibitemOpen
  \bibfield  {author} {\bibinfo {author} {\bibfnamefont {M.}~\bibnamefont {Monshizadeh}}, \bibinfo {author} {\bibfnamefont {V.}~\bibnamefont {Khatri}}, \bibinfo {author} {\bibfnamefont {R.}~\bibnamefont {Kantola}},\ and\ \bibinfo {author} {\bibfnamefont {Z.}~\bibnamefont {Yan}},\ }\bibfield  {title} {\bibinfo {title} {A deep density based and self-determining clustering approach to label unknown traffic},\ }\href {https://doi.org/https://doi.org/10.1016/j.jnca.2022.103513} {\bibfield  {journal} {\bibinfo  {journal} {Journal of Network and Computer Applications}\ }\textbf {\bibinfo {volume} {207}},\ \bibinfo {pages} {103513} (\bibinfo {year} {2022})}\BibitemShut {NoStop}%
\bibitem [{\citenamefont {Elliott}(1961)}]{elliott1961phenomenological}%
  \BibitemOpen
  \bibfield  {author} {\bibinfo {author} {\bibfnamefont {R.~J.}\ \bibnamefont {Elliott}},\ }\bibfield  {title} {\bibinfo {title} {Phenomenological discussion of magnetic ordering in the heavy rare-earth metals},\ }\href@noop {} {\bibfield  {journal} {\bibinfo  {journal} {Physical Review}\ }\textbf {\bibinfo {volume} {124}},\ \bibinfo {pages} {346} (\bibinfo {year} {1961})}\BibitemShut {NoStop}%
\bibitem [{\citenamefont {Fisher}\ and\ \citenamefont {Selke}(1980)}]{fisher1980infinitely}%
  \BibitemOpen
  \bibfield  {author} {\bibinfo {author} {\bibfnamefont {M.~E.}\ \bibnamefont {Fisher}}\ and\ \bibinfo {author} {\bibfnamefont {W.}~\bibnamefont {Selke}},\ }\bibfield  {title} {\bibinfo {title} {Infinitely many commensurate phases in a simple ising model},\ }\href@noop {} {\bibfield  {journal} {\bibinfo  {journal} {Physical Review Letters}\ }\textbf {\bibinfo {volume} {44}},\ \bibinfo {pages} {1502} (\bibinfo {year} {1980})}\BibitemShut {NoStop}%
\bibitem [{\citenamefont {Fumani}\ \emph {et~al.}(2021)\citenamefont {Fumani}, \citenamefont {Nemati},\ and\ \citenamefont {Mahdavifar}}]{fumani2021quantum}%
  \BibitemOpen
  \bibfield  {author} {\bibinfo {author} {\bibfnamefont {F.~K.}\ \bibnamefont {Fumani}}, \bibinfo {author} {\bibfnamefont {S.}~\bibnamefont {Nemati}},\ and\ \bibinfo {author} {\bibfnamefont {S.}~\bibnamefont {Mahdavifar}},\ }\bibfield  {title} {\bibinfo {title} {Quantum critical lines in the ground state phase diagram of spin-1/2 frustrated transverse-field ising chains},\ }\href@noop {} {\bibfield  {journal} {\bibinfo  {journal} {Annalen der Physik}\ }\textbf {\bibinfo {volume} {533}},\ \bibinfo {pages} {2000384} (\bibinfo {year} {2021})}\BibitemShut {NoStop}%
\bibitem [{\citenamefont {Beccaria}\ \emph {et~al.}(2007{\natexlab{b}})\citenamefont {Beccaria}, \citenamefont {Campostrini},\ and\ \citenamefont {Feo}}]{beccaria2007evidence}%
  \BibitemOpen
  \bibfield  {author} {\bibinfo {author} {\bibfnamefont {M.}~\bibnamefont {Beccaria}}, \bibinfo {author} {\bibfnamefont {M.}~\bibnamefont {Campostrini}},\ and\ \bibinfo {author} {\bibfnamefont {A.}~\bibnamefont {Feo}},\ }\bibfield  {title} {\bibinfo {title} {Evidence for a floating phase of the transverse annni model at high frustration},\ }\href@noop {} {\bibfield  {journal} {\bibinfo  {journal} {Physical Review B—Condensed Matter and Materials Physics}\ }\textbf {\bibinfo {volume} {76}},\ \bibinfo {pages} {094410} (\bibinfo {year} {2007}{\natexlab{b}})}\BibitemShut {NoStop}%
\bibitem [{\citenamefont {Chandra}\ and\ \citenamefont {Dasgupta}(2007{\natexlab{b}})}]{chandra2007floating}%
  \BibitemOpen
  \bibfield  {author} {\bibinfo {author} {\bibfnamefont {A.~K.}\ \bibnamefont {Chandra}}\ and\ \bibinfo {author} {\bibfnamefont {S.}~\bibnamefont {Dasgupta}},\ }\bibfield  {title} {\bibinfo {title} {Floating phase in the one-dimensional transverse axial next-nearest-neighbor ising model},\ }\href@noop {} {\bibfield  {journal} {\bibinfo  {journal} {Physical Review E—Statistical, Nonlinear, and Soft Matter Physics}\ }\textbf {\bibinfo {volume} {75}},\ \bibinfo {pages} {021105} (\bibinfo {year} {2007}{\natexlab{b}})}\BibitemShut {NoStop}%
\bibitem [{\citenamefont {Suzuki}\ \emph {et~al.}(2012)\citenamefont {Suzuki}, \citenamefont {Inoue},\ and\ \citenamefont {Chakrabarti}}]{suzuki2012quantum}%
  \BibitemOpen
  \bibfield  {author} {\bibinfo {author} {\bibfnamefont {S.}~\bibnamefont {Suzuki}}, \bibinfo {author} {\bibfnamefont {J.-i.}\ \bibnamefont {Inoue}},\ and\ \bibinfo {author} {\bibfnamefont {B.~K.}\ \bibnamefont {Chakrabarti}},\ }\href@noop {} {\emph {\bibinfo {title} {Quantum Ising phases and transitions in transverse Ising models}}},\ Vol.\ \bibinfo {volume} {862}\ (\bibinfo  {publisher} {Springer},\ \bibinfo {year} {2012})\BibitemShut {NoStop}%
\bibitem [{\citenamefont {Dutta}\ \emph {et~al.}(2015)\citenamefont {Dutta}, \citenamefont {Aeppli}, \citenamefont {Chakrabarti}, \citenamefont {Divakaran}, \citenamefont {Rosenbaum},\ and\ \citenamefont {Sen}}]{dutta2015quantum}%
  \BibitemOpen
  \bibfield  {author} {\bibinfo {author} {\bibfnamefont {A.}~\bibnamefont {Dutta}}, \bibinfo {author} {\bibfnamefont {G.}~\bibnamefont {Aeppli}}, \bibinfo {author} {\bibfnamefont {B.~K.}\ \bibnamefont {Chakrabarti}}, \bibinfo {author} {\bibfnamefont {U.}~\bibnamefont {Divakaran}}, \bibinfo {author} {\bibfnamefont {T.~F.}\ \bibnamefont {Rosenbaum}},\ and\ \bibinfo {author} {\bibfnamefont {D.}~\bibnamefont {Sen}},\ }\href@noop {} {\emph {\bibinfo {title} {Quantum phase transitions in transverse field spin models: from statistical physics to quantum information}}}\ (\bibinfo  {publisher} {Cambridge University Press},\ \bibinfo {year} {2015})\BibitemShut {NoStop}%
\bibitem [{\citenamefont {Nagy}(2011{\natexlab{b}})}]{nagy2011exploring}%
  \BibitemOpen
  \bibfield  {author} {\bibinfo {author} {\bibfnamefont {A.}~\bibnamefont {Nagy}},\ }\bibfield  {title} {\bibinfo {title} {Exploring phase transitions by finite-entanglement scaling of mps in the 1d annni model},\ }\href@noop {} {\bibfield  {journal} {\bibinfo  {journal} {New Journal of Physics}\ }\textbf {\bibinfo {volume} {13}},\ \bibinfo {pages} {023015} (\bibinfo {year} {2011}{\natexlab{b}})}\BibitemShut {NoStop}%
\bibitem [{\citenamefont {Bonfim}\ \emph {et~al.}(2017)\citenamefont {Bonfim}, \citenamefont {Boechat},\ and\ \citenamefont {Florencio}}]{bonfim2017quantum}%
  \BibitemOpen
  \bibfield  {author} {\bibinfo {author} {\bibfnamefont {O.~d.~A.}\ \bibnamefont {Bonfim}}, \bibinfo {author} {\bibfnamefont {B.}~\bibnamefont {Boechat}},\ and\ \bibinfo {author} {\bibfnamefont {J.}~\bibnamefont {Florencio}},\ }\bibfield  {title} {\bibinfo {title} {Quantum fidelity approach to the ground-state properties of the one-dimensional axial next-nearest-neighbor ising model in a transverse field},\ }\href@noop {} {\bibfield  {journal} {\bibinfo  {journal} {Physical Review E}\ }\textbf {\bibinfo {volume} {96}},\ \bibinfo {pages} {042140} (\bibinfo {year} {2017})}\BibitemShut {NoStop}%
\bibitem [{\citenamefont {Nemati}\ \emph {et~al.}(2020)\citenamefont {Nemati}, \citenamefont {Khastehdel~Fumani},\ and\ \citenamefont {Mahdavifar}}]{nemati2020comment}%
  \BibitemOpen
  \bibfield  {author} {\bibinfo {author} {\bibfnamefont {S.}~\bibnamefont {Nemati}}, \bibinfo {author} {\bibfnamefont {F.}~\bibnamefont {Khastehdel~Fumani}},\ and\ \bibinfo {author} {\bibfnamefont {S.}~\bibnamefont {Mahdavifar}},\ }\bibfield  {title} {\bibinfo {title} {Comment on “quantum fidelity approach to the ground-state properties of the one-dimensional axial next-nearest-neighbor ising model in a transverse field”},\ }\href@noop {} {\bibfield  {journal} {\bibinfo  {journal} {Physical Review E}\ }\textbf {\bibinfo {volume} {102}},\ \bibinfo {pages} {016101} (\bibinfo {year} {2020})}\BibitemShut {NoStop}%
\bibitem [{\citenamefont {Sen}\ \emph {et~al.}(1992)\citenamefont {Sen}, \citenamefont {Chakraborty}, \citenamefont {Dasgupta},\ and\ \citenamefont {Chakrabarti}}]{sen1992numerical}%
  \BibitemOpen
  \bibfield  {author} {\bibinfo {author} {\bibfnamefont {P.}~\bibnamefont {Sen}}, \bibinfo {author} {\bibfnamefont {S.}~\bibnamefont {Chakraborty}}, \bibinfo {author} {\bibfnamefont {S.}~\bibnamefont {Dasgupta}},\ and\ \bibinfo {author} {\bibfnamefont {B.}~\bibnamefont {Chakrabarti}},\ }\bibfield  {title} {\bibinfo {title} {Numerical estimate of the phase diagram of finite annni chains in transverse field},\ }\href@noop {} {\bibfield  {journal} {\bibinfo  {journal} {Zeitschrift f{\"u}r Physik B Condensed Matter}\ }\textbf {\bibinfo {volume} {88}},\ \bibinfo {pages} {333} (\bibinfo {year} {1992})}\BibitemShut {NoStop}%
\bibitem [{\citenamefont {Gray}(2018)}]{gray2018quimb}%
  \BibitemOpen
  \bibfield  {author} {\bibinfo {author} {\bibfnamefont {J.}~\bibnamefont {Gray}},\ }\bibfield  {title} {\bibinfo {title} {quimb: A python package for quantum information and many-body calculations},\ }\href@noop {} {\bibfield  {journal} {\bibinfo  {journal} {Journal of Open Source Software}\ }\textbf {\bibinfo {volume} {3}},\ \bibinfo {pages} {819} (\bibinfo {year} {2018})}\BibitemShut {NoStop}%
\bibitem [{\citenamefont {Eisert}\ \emph {et~al.}(2010)\citenamefont {Eisert}, \citenamefont {Cramer},\ and\ \citenamefont {Plenio}}]{eisert2010colloquium}%
  \BibitemOpen
  \bibfield  {author} {\bibinfo {author} {\bibfnamefont {J.}~\bibnamefont {Eisert}}, \bibinfo {author} {\bibfnamefont {M.}~\bibnamefont {Cramer}},\ and\ \bibinfo {author} {\bibfnamefont {M.~B.}\ \bibnamefont {Plenio}},\ }\bibfield  {title} {\bibinfo {title} {Colloquium: Area laws for the entanglement entropy},\ }\href@noop {} {\bibfield  {journal} {\bibinfo  {journal} {Reviews of modern physics}\ }\textbf {\bibinfo {volume} {82}},\ \bibinfo {pages} {277} (\bibinfo {year} {2010})}\BibitemShut {NoStop}%
\bibitem [{\citenamefont {Ohta}\ \emph {et~al.}(2016)\citenamefont {Ohta}, \citenamefont {Tanaka}, \citenamefont {Danshita},\ and\ \citenamefont {Totsuka}}]{ohta2016topological}%
  \BibitemOpen
  \bibfield  {author} {\bibinfo {author} {\bibfnamefont {T.}~\bibnamefont {Ohta}}, \bibinfo {author} {\bibfnamefont {S.}~\bibnamefont {Tanaka}}, \bibinfo {author} {\bibfnamefont {I.}~\bibnamefont {Danshita}},\ and\ \bibinfo {author} {\bibfnamefont {K.}~\bibnamefont {Totsuka}},\ }\bibfield  {title} {\bibinfo {title} {Topological and dynamical properties of a generalized cluster model in one dimension},\ }\href@noop {} {\bibfield  {journal} {\bibinfo  {journal} {Physical Review B}\ }\textbf {\bibinfo {volume} {93}},\ \bibinfo {pages} {165423} (\bibinfo {year} {2016})}\BibitemShut {NoStop}%
\bibitem [{\citenamefont {Son}\ \emph {et~al.}(2011)\citenamefont {Son}, \citenamefont {Amico}, \citenamefont {Fazio}, \citenamefont {Hamma}, \citenamefont {Pascazio},\ and\ \citenamefont {Vedral}}]{son2011quantum}%
  \BibitemOpen
  \bibfield  {author} {\bibinfo {author} {\bibfnamefont {W.}~\bibnamefont {Son}}, \bibinfo {author} {\bibfnamefont {L.}~\bibnamefont {Amico}}, \bibinfo {author} {\bibfnamefont {R.}~\bibnamefont {Fazio}}, \bibinfo {author} {\bibfnamefont {A.}~\bibnamefont {Hamma}}, \bibinfo {author} {\bibfnamefont {S.}~\bibnamefont {Pascazio}},\ and\ \bibinfo {author} {\bibfnamefont {V.}~\bibnamefont {Vedral}},\ }\bibfield  {title} {\bibinfo {title} {Quantum phase transition between cluster and antiferromagnetic states},\ }\href@noop {} {\bibfield  {journal} {\bibinfo  {journal} {Europhysics Letters}\ }\textbf {\bibinfo {volume} {95}},\ \bibinfo {pages} {50001} (\bibinfo {year} {2011})}\BibitemShut {NoStop}%
\bibitem [{\citenamefont {Smacchia}\ \emph {et~al.}(2011)\citenamefont {Smacchia}, \citenamefont {Amico}, \citenamefont {Facchi}, \citenamefont {Fazio}, \citenamefont {Florio}, \citenamefont {Pascazio},\ and\ \citenamefont {Vedral}}]{smacchia2011statistical}%
  \BibitemOpen
  \bibfield  {author} {\bibinfo {author} {\bibfnamefont {P.}~\bibnamefont {Smacchia}}, \bibinfo {author} {\bibfnamefont {L.}~\bibnamefont {Amico}}, \bibinfo {author} {\bibfnamefont {P.}~\bibnamefont {Facchi}}, \bibinfo {author} {\bibfnamefont {R.}~\bibnamefont {Fazio}}, \bibinfo {author} {\bibfnamefont {G.}~\bibnamefont {Florio}}, \bibinfo {author} {\bibfnamefont {S.}~\bibnamefont {Pascazio}},\ and\ \bibinfo {author} {\bibfnamefont {V.}~\bibnamefont {Vedral}},\ }\bibfield  {title} {\bibinfo {title} {Statistical mechanics of the cluster ising model},\ }\href@noop {} {\bibfield  {journal} {\bibinfo  {journal} {Physical Review A—Atomic, Molecular, and Optical Physics}\ }\textbf {\bibinfo {volume} {84}},\ \bibinfo {pages} {022304} (\bibinfo {year} {2011})}\BibitemShut {NoStop}%
\bibitem [{\citenamefont {Skr{\o}vseth}\ and\ \citenamefont {Bartlett}(2009)}]{skrovseth2009phase}%
  \BibitemOpen
  \bibfield  {author} {\bibinfo {author} {\bibfnamefont {S.~O.}\ \bibnamefont {Skr{\o}vseth}}\ and\ \bibinfo {author} {\bibfnamefont {S.~D.}\ \bibnamefont {Bartlett}},\ }\bibfield  {title} {\bibinfo {title} {Phase transitions and localizable entanglement in cluster-state spin chains with ising couplings and local fields},\ }\href@noop {} {\bibfield  {journal} {\bibinfo  {journal} {Physical Review A—Atomic, Molecular, and Optical Physics}\ }\textbf {\bibinfo {volume} {80}},\ \bibinfo {pages} {022316} (\bibinfo {year} {2009})}\BibitemShut {NoStop}%
\bibitem [{\citenamefont {Strinati}\ \emph {et~al.}(2017)\citenamefont {Strinati}, \citenamefont {Rossini}, \citenamefont {Fazio},\ and\ \citenamefont {Russomanno}}]{strinati2017resilience}%
  \BibitemOpen
  \bibfield  {author} {\bibinfo {author} {\bibfnamefont {M.~C.}\ \bibnamefont {Strinati}}, \bibinfo {author} {\bibfnamefont {D.}~\bibnamefont {Rossini}}, \bibinfo {author} {\bibfnamefont {R.}~\bibnamefont {Fazio}},\ and\ \bibinfo {author} {\bibfnamefont {A.}~\bibnamefont {Russomanno}},\ }\bibfield  {title} {\bibinfo {title} {Resilience of hidden order to symmetry-preserving disorder},\ }\href@noop {} {\bibfield  {journal} {\bibinfo  {journal} {Physical Review B}\ }\textbf {\bibinfo {volume} {96}},\ \bibinfo {pages} {214206} (\bibinfo {year} {2017})}\BibitemShut {NoStop}%
\bibitem [{\citenamefont {Herrmann}\ \emph {et~al.}(2022)\citenamefont {Herrmann}, \citenamefont {Llima}, \citenamefont {Remm}, \citenamefont {Zapletal}, \citenamefont {McMahon}, \citenamefont {Scarato}, \citenamefont {Swiadek}, \citenamefont {Andersen}, \citenamefont {Hellings}, \citenamefont {Krinner} \emph {et~al.}}]{herrmann2022realizing}%
  \BibitemOpen
  \bibfield  {author} {\bibinfo {author} {\bibfnamefont {J.}~\bibnamefont {Herrmann}}, \bibinfo {author} {\bibfnamefont {S.~M.}\ \bibnamefont {Llima}}, \bibinfo {author} {\bibfnamefont {A.}~\bibnamefont {Remm}}, \bibinfo {author} {\bibfnamefont {P.}~\bibnamefont {Zapletal}}, \bibinfo {author} {\bibfnamefont {N.~A.}\ \bibnamefont {McMahon}}, \bibinfo {author} {\bibfnamefont {C.}~\bibnamefont {Scarato}}, \bibinfo {author} {\bibfnamefont {F.}~\bibnamefont {Swiadek}}, \bibinfo {author} {\bibfnamefont {C.~K.}\ \bibnamefont {Andersen}}, \bibinfo {author} {\bibfnamefont {C.}~\bibnamefont {Hellings}}, \bibinfo {author} {\bibfnamefont {S.}~\bibnamefont {Krinner}}, \emph {et~al.},\ }\bibfield  {title} {\bibinfo {title} {Realizing quantum convolutional neural networks on a superconducting quantum processor to recognize quantum phases},\ }\href@noop {} {\bibfield  {journal} {\bibinfo  {journal} {Nature communications}\ }\textbf {\bibinfo {volume} {13}},\ \bibinfo {pages} {4144} (\bibinfo {year} {2022})}\BibitemShut
  {NoStop}%
\bibitem [{\citenamefont {Zhou}\ \emph {et~al.}(2017)\citenamefont {Zhou}, \citenamefont {Kanoda},\ and\ \citenamefont {Ng}}]{zhou2017quantum}%
  \BibitemOpen
  \bibfield  {author} {\bibinfo {author} {\bibfnamefont {Y.}~\bibnamefont {Zhou}}, \bibinfo {author} {\bibfnamefont {K.}~\bibnamefont {Kanoda}},\ and\ \bibinfo {author} {\bibfnamefont {T.-K.}\ \bibnamefont {Ng}},\ }\bibfield  {title} {\bibinfo {title} {Quantum spin liquid states},\ }\href@noop {} {\bibfield  {journal} {\bibinfo  {journal} {Reviews of Modern Physics}\ }\textbf {\bibinfo {volume} {89}},\ \bibinfo {pages} {025003} (\bibinfo {year} {2017})}\BibitemShut {NoStop}%
\bibitem [{\citenamefont {Wen}\ \emph {et~al.}(2019)\citenamefont {Wen}, \citenamefont {Yu}, \citenamefont {Li}, \citenamefont {Yu},\ and\ \citenamefont {Li}}]{wen2019experimental}%
  \BibitemOpen
  \bibfield  {author} {\bibinfo {author} {\bibfnamefont {J.}~\bibnamefont {Wen}}, \bibinfo {author} {\bibfnamefont {S.-L.}\ \bibnamefont {Yu}}, \bibinfo {author} {\bibfnamefont {S.}~\bibnamefont {Li}}, \bibinfo {author} {\bibfnamefont {W.}~\bibnamefont {Yu}},\ and\ \bibinfo {author} {\bibfnamefont {J.-X.}\ \bibnamefont {Li}},\ }\bibfield  {title} {\bibinfo {title} {Experimental identification of quantum spin liquids},\ }\href@noop {} {\bibfield  {journal} {\bibinfo  {journal} {npj Quantum Materials}\ }\textbf {\bibinfo {volume} {4}},\ \bibinfo {pages} {12} (\bibinfo {year} {2019})}\BibitemShut {NoStop}%
\bibitem [{\citenamefont {Binder}\ and\ \citenamefont {Young}(1986)}]{binder1986spin}%
  \BibitemOpen
  \bibfield  {author} {\bibinfo {author} {\bibfnamefont {K.}~\bibnamefont {Binder}}\ and\ \bibinfo {author} {\bibfnamefont {A.~P.}\ \bibnamefont {Young}},\ }\bibfield  {title} {\bibinfo {title} {Spin glasses: Experimental facts, theoretical concepts, and open questions},\ }\href@noop {} {\bibfield  {journal} {\bibinfo  {journal} {Reviews of Modern physics}\ }\textbf {\bibinfo {volume} {58}},\ \bibinfo {pages} {801} (\bibinfo {year} {1986})}\BibitemShut {NoStop}%
\bibitem [{\citenamefont {Huang}\ \emph {et~al.}(2020)\citenamefont {Huang}, \citenamefont {Kueng},\ and\ \citenamefont {Preskill}}]{huang2020predicting}%
  \BibitemOpen
  \bibfield  {author} {\bibinfo {author} {\bibfnamefont {H.-Y.}\ \bibnamefont {Huang}}, \bibinfo {author} {\bibfnamefont {R.}~\bibnamefont {Kueng}},\ and\ \bibinfo {author} {\bibfnamefont {J.}~\bibnamefont {Preskill}},\ }\bibfield  {title} {\bibinfo {title} {Predicting many properties of a quantum system from very few measurements},\ }\href@noop {} {\bibfield  {journal} {\bibinfo  {journal} {Nature Physics}\ }\textbf {\bibinfo {volume} {16}},\ \bibinfo {pages} {1050} (\bibinfo {year} {2020})}\BibitemShut {NoStop}%
\bibitem [{\citenamefont {Cie{\'s}li{\'n}ski}\ \emph {et~al.}(2024)\citenamefont {Cie{\'s}li{\'n}ski}, \citenamefont {Imai}, \citenamefont {Dziewior}, \citenamefont {G{\"u}hne}, \citenamefont {Knips}, \citenamefont {Laskowski}, \citenamefont {Meinecke}, \citenamefont {Paterek},\ and\ \citenamefont {V{\'e}rtesi}}]{cieslinski2024analysing}%
  \BibitemOpen
  \bibfield  {author} {\bibinfo {author} {\bibfnamefont {P.}~\bibnamefont {Cie{\'s}li{\'n}ski}}, \bibinfo {author} {\bibfnamefont {S.}~\bibnamefont {Imai}}, \bibinfo {author} {\bibfnamefont {J.}~\bibnamefont {Dziewior}}, \bibinfo {author} {\bibfnamefont {O.}~\bibnamefont {G{\"u}hne}}, \bibinfo {author} {\bibfnamefont {L.}~\bibnamefont {Knips}}, \bibinfo {author} {\bibfnamefont {W.}~\bibnamefont {Laskowski}}, \bibinfo {author} {\bibfnamefont {J.}~\bibnamefont {Meinecke}}, \bibinfo {author} {\bibfnamefont {T.}~\bibnamefont {Paterek}},\ and\ \bibinfo {author} {\bibfnamefont {T.}~\bibnamefont {V{\'e}rtesi}},\ }\bibfield  {title} {\bibinfo {title} {Analysing quantum systems with randomised measurements},\ }\href@noop {} {\bibfield  {journal} {\bibinfo  {journal} {Physics Reports}\ }\textbf {\bibinfo {volume} {1095}},\ \bibinfo {pages} {1} (\bibinfo {year} {2024})}\BibitemShut {NoStop}%
\end{thebibliography}%

\end{document}